\documentclass[]{article}

\usepackage{graphicx}
\usepackage{dcolumn}
\usepackage{bm}
\usepackage[utf8]{inputenc}
\usepackage[T1]{fontenc}
\usepackage{mathptmx}
\usepackage{amsfonts}
\usepackage{bbm}
\usepackage[bottom]{footmisc}
\usepackage[utf8]{inputenc}
\usepackage{amsmath}
\usepackage[svgnames]{xcolor}
\usepackage[colorlinks=true, allcolors=DarkBlue]{hyperref}

\begin{document}

\title{An alternative foundation of quantum theory}

\author{Inge S. Helland \\
              Department of Mathematics, University of Oslo\\ P.O. Box 1053, 0316 Oslo, Norway \\
              Tel.: +47-93688918\\
            ingeh@math.uio.no}

\date{}

\maketitle

\begin{abstract}
The final version of a new approach to quantum theory is formulated in this paper. The basis is taken to be theoretical variables, variables that may be accessible or inaccessible, i.e., it may be possible or impossible for an observer to assign arbitrarily sharp numerical values to them. In an epistemic process, the accessible variables are just ideal observations connected to an observer or to some communicating observers. Group actions are defined on these variables, and group representation theory is the basis for developing the Hilbert space formalism here. Operators corresponding to accessible theoretical variables are derived, and in the finite-dimensional case, it is proved that the possible physical values are the eigenvalues of these operators. The focus of the paper is some mathematical theorems paving the ground for the proposed foundation of quantum theory. It is indicated here that the groups and transformations needed in this approach can be constructed explicitly in the case where the accessible variables are finite-dimensional. In case, this simplifies the theory considerably: To reproduce the Hilbert space formulation, it is enough to assume the existence of two complementary variables. What is new in this version of the article, is that the assumptions given behind the main theorems are simplified considerably, and that all proofs are given explicitly or refered to in other recent papers.

\end{abstract}

\underline{Keywords:} Accessible theoretical variables; group theory; group representation theory; partial ordering; quantum operators; quantum states; quantum foundation.

\clearpage
\thispagestyle{empty}
\begin{quote}
\tableofcontents
\end{quote}
\thispagestyle{empty}
\clearpage
\setcounter{page}{1}

\section{Introduction}
\label{intro}

This article  is a new version of the published article [81], with the differences that weakenings of the mathematical assumptions are made, using in an essential way the results given in the recent article [82]. The conclusions derived by the theory are the same, but these changes make the theory simpler and more general.

For an outsider, one of the really difficult things to accept about quantum mechanics is its state concept: The state of a physical system is given by a normalized vector in a complex separable Hilbert space. One question that I will raise here, is whether this state concept can be derived, or at least motivated, by some other considerations. The discussions given here, may be seen as a continuation of the discussions in the book [1], and a further discussion of the foundation advocated in [81].

My point of departure will still be the notion of theoretical variables, variables attached to some observer or to a group of communicating observers. This will be taken as a very wide concept, but it includes physical variables like position, time, momentum and energy. The term `theoretical' points to a distinction from real, measured variables, that in a typical connection in the continuous case may be seen as a theoretical variable plus some random error.

 The notion of theoretical variables can also be connected to several interpretations of quantum mechanics.
To give an example of such a connection, according to Relational Quantum Mechanics, see Rovelli [2] and van Fraassen [3], variables take values only at interactions, and the values that they take are only relative to the other system affected by the interaction. This other system might well be an observer, and I will think of such a situation. It is crucial to me that certain physical phenomena only may be explained be refering to what is seen by an observer or by a communicating group of observers.

 Other examples of theoretical variables are decision variables in a quantum decision context. This will be briefly discussed below, and is discussed in more details in [5,83]. 

Variables which can take definite values relative to an observer or group of observers will be called accessible. But relative to the observers, there may also be other theoretical variables which I will call inaccessible. An example may be the vector (position, momentum) connected to the observation of a particle. Another example may be the spin vector of a particle, where I think of a mental model of the spin, where the spin components are seen as quantizised values of the projection of this vector upon some given direction.

These inaccessible variables must be seen as purely mathematical variables; from a physical point of view, they do not take any known values, but nevertheless they can be seen as variables. 

The distinction between accessible and inaccessible variables is very important to my approach. On the one hand, this distinction can be explained roughly to outsiders. On the other hand, one can give many precise physical examples. In the book  [84] with Harish Parthasarathy, we among other things illustrate this distinction on several examples from quantum field theory.

In this article, I will discuss these notions more closely with a focus on the more mathematical aspects of the situations described above. First, concentrate on the variables and the corresponding group theory. I assume the existence of a concrete (physical) situation, and that there is a space $\Omega_\phi$ on which an inaccessible theoretical variable $\phi$ varies, with a group $K$ acting on this space. There is at least one accessible theoretical variable, $\theta$, defined, where it is assumed that it can be seen as a function on $\Omega_\phi$. This is a crucial assumption.

This $\theta$ varies on a space $\Omega_\theta$, and the group $K$ may or may not induce a transformation group $G$ on $\Omega_\theta$. In any case, I will focus on such a group $G$ on $\Omega_\theta$, whether it is induced by $K$ or not. An essential requirement is that $G$ is transitive on $\Omega_\theta$. It is shown in [81], and repeated below, that the existence of $G$, together with symmetry assumptions assumed in [4], will be satisfied without further assumptions in the finite-dimensional case.

A special situation is when $\phi$ is a spin vector, and $\theta$ is a spin component in a given direction. In the simple spin situation, the natural group $K$ for the spin vector does not directly induce groups on the components. But does so if we redefine $\phi$ to be the projection of the spin vector upon the plane spanned by two different directions in which spin can be measured and take $K$ to be the corresponding rotation group.

When there are several potential accessible variables, I may denote this by a superscript $a$: $\theta^a$ for the variables and $G^a$ for the groups, with elements $g^a$. Both here and in [1,4,5, 81, 82] I use the word `group' as synonymous to `group action' or transformation group on some set, not as an abstract group.

This article is an extended version of the paper [81], and it also has some overlap with the papers [4,5,82], but the results there are further discussed and clarified here. 

My derivations here will be independent of other Hilbert space reconstructions in the literature, but they will in some sense compete with several rather deep investigations recently on deriving the Hilbert space structure from various assumptions [6,7,8,9,10]. By relying on group representation theory, I use at the outset some Hilbert space structure, but this is by construction, not by assumption. The construction is indicated to be realizable in the important finite-dimensional case. It is interesting to see, as stated in [11], that there is a problem connecting the above general derivations to the many different interpretations of quantum theory. By contrast, the derivation presented here seems to lead naturally to a particular interpretation: A general epistemic interpretation, which also may have links to an ontological interpretation in many cases, see below.

There is a large literature on different interpretations of quantum mechanics. In my opinion, one should try to clarify the questions around quantum foundations before one goes into a deeper discussion of possible interpretations. Brief discussions of interpretation are given in a later Section of this article. This discussion is in agreement with the view expressed by Robert Spekkens in a recent video: It may be seen as a categorical error to look upon quantum theory as describing the real world; it is a theory of our \emph{knowledge} of the world. Of course the word knowledge can be interpreted in many ways, and an epistemic view upon quantum mechanics may be discussed, but in my opinion a focus on describing our knowledge of the world leads to more attractive interpretations of the theory.

There has been many attempts in the literature to motivate the Hilbert space formalism. What is special about my approach? First, it is based upon notions that can be communicated to outsiders with a limited mathematical background. Secondly, it is based on very few postulates. One postulate, Postulate 3 in Section 3 below, may be seen by some people as strong, but this postulate is discussed here from several points of view. One view is a possible relation between science and religion; see also [77,80].

Finally, I see it as crucial that my main result, also formulated in Section 3, in addition to the postulates only assumes the existence of just two different maximal accessible variables in order to establish essential elements of the Hilbert space apparatus, These two variables may be seen, in the language of Niels Bohr, to be complementary.

One of the last derivations of the Hilbert space formalism is Brezhnev [78], where also other derivations are referred to. [78] is based upon an unlimited validity of the superposition principle. As a contrast, in the present approach the operators associated with theoretical variables are derived first, and a ket vector is only seen as a valid state vector if it is an eigenvector of a physically meaningful operator. This leads to a limitation of the superposition principle, but in my opinion, it facilitates the interpretation of quantum mechanics.

Several applications of the theory given here can be mentioned. One application is to give an explanation [34,61] of the fact that Bell's inequalities may be violated in real experiments, contrary to what could be expected intuitively. Other applications are discussions of the two-slit experiment and the paradoxes connected to Schr\"{o}dinger's cat and Wigner's friend [47]. In [47] and [84], we also consider links to field theory and to general relativity theory.

The link to quantum decision theory [5,83] is under further considerations. One goal is to find connections to statistical inference theory, where my accessible theoretical variables will be statistical parameters. (I have avoided the term `parameter' in this article, since in physics this word has other meanings.) This is now to some extent achieved; see [86, 87]. Such interdisciplinary investigations would have been impossible with the common quantum theory formalism.

In my proofs I use group representation theory in an essential way. Group representation theory in discussing quantum foundation has also been used in other places; see for instance [12]. In quantum field theory and particle physics theory, the use of group representation theory is crucial [13, 79].

Sections 2 and 3 give some background. In Sections 4 and 5 I introduce some basic group theory and group representation theory that is needed in the paper. Then, in Sections 6 and 7, I formulate my approach to the foundation of quantum theory. Simple postulates for the relevant situations are assumed. From this, the ordinary quantum formalism is derived, and it is shown how operators attached to accessible physical variables may be defined. A brief comparison with some other approaches towards quantum foundation is included in Section 8. In Section 9 a corresponding interpretation of quantum mechanics is discussed, in Section 10 a slightly different approach based upon category thory is discussed, and Section 11 gives some concluding remarks.

To complete the derivation of quantum theory along these lines, one will also need a derivation of the Born rule under suitable conditions, and a derivation of the Schr\"{o}dinger equation. A very brief discussion of this is included in the crucial Section 7 for completeness; a more thorough discussion is given in [1, 88]. The main focus of the present article is the construction of the Hilbert space apparatus from simple assumptions and the corresponding interpretation. It must also be emphasized that there are many alternative derivetions of the Born rule from different assumptions in the literature. For instance, in a recent preprint [85], Zayn Blore derives the rule from two symmetry assumptions.

\section{Convivial Solipsism}

In Herv\'{e} Zwirn's paper [30] it is shown how a long range of foundational issues in quantum mechanics, including the famous measurement problem, can be enlightened using a new philosophy called Convivial Solipsism. The basic thesis is: \emph{Every description of the world must be relative to some observer. But different observers can communicate.} Mathematically, the philosophy rests upon Everett's relative state formulation of pure wave mechanics [49, 53]; see also [71] for philosophical issues. The distinction between absolute states, which can be taken to describe the whole, including physical system, measuring apparatus and observer, on the one hand, and relative states, describing, say, the state of an observer's brain after a measurement, is crucial.

In concrete terms, assume a system which can be in one out of two states $|S1\rangle$ or $|S2\rangle$; the corresponding states of the measurement apparatus are $|E1\rangle$ and $E2\rangle$, while this induces states of the observer's brain $|B1\rangle$, respectively $|B2\rangle$. All these are relative states. The global state will be of the form
\begin{equation}
|\psi\rangle = \alpha |S1\rangle |E1\rangle |B1\rangle +\beta |S2\rangle |E2\rangle |B2\rangle ),
\label{globalstate}
\end{equation}

Zwirn distinguishes between the states of the brain and the `states' of the consciousness of the corresponding observer, which can be denoted by $\tilde{B1}$ or $\tilde{B2}$. From this, the measurement problem and other problems related to the interpretation of quantum mechanics are discussed. The basic principles behind this discussion are the hanging-up mechanism and the relativity of states assumption; see [30].

All this presupposes a foundation of quantum mechanics based upon ket vectors as representing physical states. Below, I will try to avoid this formal apparatus as a starting point and start with completely different notions. But I will keep Zwirn's basic thesis, with the possible addition that I will allow the notion of an observer to be replaced by a group of communicating observers, where these observers are assumed to be able to communicate about everything related to the relevant theoretical variables.

\section{Motivation, first postulates, and some basic results}

Can one find a new foundation of quantum theory, a foundation that ultimately leads to the full theory, but at the same time a foundation that can also be explained to persons who never have been exposed to the ordinary Hilbert space machinery?

My answer is yes. I have up to now discussed my approach in four books [1,24, 83, 84] and in several papers [4,5,25,34,47,61,73,81,82,86,87,88,90]. Now I aim to collect the mathematical arguments in a single article and also give some results beyond the above books and articles.

My basic notion is that of a theoretical variable, which is a very wide notion. This variable can be a physical variable, a statistical parameter, a future data variable, a decision variable, or perhaps also other things. In this discussion, the variables can always be seen as mathematical variables. I divide the variables into accessible ones and inaccessible ones, as briefly discussed in the Introduction. From a mathematical point of view, I only require that if $\theta$ is accessible and $\lambda$ is a variable which is a function of $\theta$, then $\lambda$ is also accessible. 

Here are some examples:

1) Spin of one particle. An observer $C$ can have the choice between measuring the spin component in the $x$ direction or in the $z$ direction. This gives two different accessible variables related to $C$. An inaccessible variable is the unit spin vector $\phi$, which we think of as a three dimensional vector such that the spin component in a certain direction is a discretized version of the projection of $\phi$ in that direction. In the qubit case, the spin component in any direction $a$ can be modeled as a simple function of $\phi$: $\theta^a = \mathrm{sign}(\mathrm{cos}(a,\phi))$, taking the values -1 or +1. A correct distribution of each $\theta^a$ will result if we let $\phi$ have a uniform distribution on the unit sphere. By using the Born rule, which requires a separate derivation, a conditional distribution of $\theta^b$, given $\theta^a$, $a\ne b$, can also be found. This is the first example of quantum mechanics discussed in the book [48].

2) The EPR situation (as modified by Bohm) with actors Alice and Bob. For an independent observer Charlie, the unit spin vectors of both are inaccessible, say $\phi_A$ and $\phi_B$. But it can be shown that the dot product of the two is accessible to Charlie: $\eta =\phi_A \cdot \phi_B$. Specifically, one can show that Charlie is associated with an eigenstate for the operator corresponding to the variable $\eta$, which is the entangled singlet state corresponding to $\eta=-3$. One can show that this implies that for Charlie and for the measured components in some fixed direction $a$, the component of Alice is opposite to the component of Bob. Note that Charlie can be any person. See Subsection 7.5 below for more details.

3) The Bell experiment situation. Look at the subsample of data where Alice measures her spin component in direction $a$ and gets a response $A$, either -1 or +1, and where Bob measures in a direction $b$ and gets a similar response $B$. For an independent observer Charlie, analysing the data after the experiment, both $A$ and $B$ are accessible. This implies by Born's formula – anticipating this formula - a fixed joint distribution of $A$ and $B$. But Charlie has his limitations, as in 2). In an article [34] in Foundations of Physics, I discuss what this limitation implies for him, using my point of view. This may be used to explain the now well-known violation of Bell's inequality in practice. Unfortunately [34] contains some smaller errors, which were corrected later in the article [61].

4) Consider first a general decision problem with two alternatives. In the simplest case the observer $C$ knows the consequences of both choices, they can both be seen as results given by an accessible variable. But in more complicated cases, the consequences are inaccessible, and hence the results of his choice are inaccessible. Then an option can be to make a simpler sub-decision, where he knows the consequences. A decision variable can be connected to a decision process, and this decision variable can be said to be accessible if $C$ is able to make a decision. Maximal accessible decision variables seem to be of some interest here. Quantum decision problems are discussed in [5], where further references are given. See also Subsection 7.4 below.

All these examples can be coupled to my approach towards quantum mechanics. I will now sketch the basic elements of this approach.

My point of departure is a statement of Hervé Zwirn’s Convivial Solipsism, as noted before: Every description of the world must be relative to some observer $C$ or to a group of communicating observers. I will assume that this observer/ these observers is/ are in some fixed physical or non-physical  context. My primitive concept is then theoretical variables related to this situation.

A concrete area where a non-physical context is meaningful and required is quantum decision theory; see Subsection 7.4.

At the outset, I assume that there is a partial ordering among the theoretical variables.  I define a concrete partial ordering based on functional dependence. I thus only assume
\smallskip

\textbf{Postulate 1:} \textit{ If $\theta$ is a theoretical variable, and $\eta = f(\theta)$ for some function $f$, then $\eta$ is a theoretical variable.}
\bigskip

The theoretical variables may be accessible or inaccessible to $C$ (or to the group).Very roughly we can say: If $\theta$ is accessible, $C$ will, in principle in some future be able to find an accurate value of $\theta$ as he likes. But as a referee remarks, this rough definition raises many questions: What is meant by the future? Is the accuracy limited by the Planck length? - and so on.

So I will just take `accessible' as a primitive notion. From a mathematical point of view, I only
assume:
\bigskip

\textbf{Postulate 2:} \textit{If $\theta$ is accessible to $C$ and $\lambda = f(\theta)$ for some function $f$ , then $\lambda$ is also accessible to $C$.}
\bigskip

Depending upon the situation, the choice of classes of functions $f$ may be different. In some applications, the variables may be scalars or vectors, and we may consider just linear functions, in other applications, we may have topologies on the range spaces $\Omega_\theta$, $\Omega_\eta$, and $\Omega_\lambda$, and consider all Borel-measurable functions.

The crucial model assumption is now the following (; see also [1,4,5]):
\bigskip

\textbf{Postulate 3:} \textit{In the given context there exists an inaccessible variable $\phi$ such that
all the accessible variables can be seen as functions of $\phi$. There is a group $K$ acting on the range $\Omega_\phi$ of the variable $\phi$.}
\bigskip

First look at two physical examples:

A) Consider a Stern-Gerlach experiment where the spin component of a particle is measured in an arbitrary direction $a$. These spin components are accessible variables, and can, as noted in 1) above be seen as functions of an inaccessible variable, the full spin vector, which only exists theoretically relative to the relevant observer.

B) The position of a particle is accessible, and so is its momentum. Both are functions of the variable $\phi$=(position, momentum), which is inaccessible by Heisenberg's inequality.

In any more general mathematical setting, following ideas by Palmer [76], Postulate 3 can be motivated as follows in the finite-dimensional case, say where maximal accessible variables have dimension $n$. (See Definition 1 below for a definition of `maximal'): Assume that the variable $\phi$ is generated by some chaotic process, and let it be written in base-$n$. Then we can let the maximal accessible variable $\theta$ be determined by the millonth digit in $\phi$, the maximal variable $\eta$ be determined by digit number 1000 003, and so on. Then by Postulate 4 below, all accessible variables can be seen as some function of a maximal accessible variable, so Postulate 3 will hold in this setting. Note, however, that with this choice of $\phi$, it is difficult to define a transitive group $K$ on $\Omega_\phi$ such that the functions $\theta(\cdot)$ are permissible with respect to $K$ (see later).

A perhaps more satisfying background for Postulate 3, is to assume some relation between science and religion. I know that many scientists are skeptical to such relationships, but here it seems to me to clarify the possible basis of quantum theory. One can simply assume that $\phi$ is known to God, but unknown to us humans. This also explains why one can think of a quantum foundation attached to a human observer. Presumably, quantum mechanics and its generalizations have existed in some form forever, but human beings only for some million of years. 

As a background for my theory, I assume that God has existed forever, and so have basic physical laws. A rather common assumption in various religions is that we humans are created in God's image. Thus, in very metaphysical terms, one can simplify a theory of God's mind by a theory of an observer's mind, which is partly done in this article. For more on my views on religion and quantum foundation, see [77, 80].

In general, it should only be noted that $\phi$ is a variable and is not assumed to take any value known to us. This is also analoguous to the main assumptions of probability theory, which is based upon functions on a probability space $(\Omega, \mathcal{F},P)$, with no assumption that $\omega\in\Omega$ is known to us.

It will be shown below that Postulate 3, taken together with some symmetry assumptions, has far-reaching consequences. It will imply, given the existence of two different maximal accessible variables, the whole Hilbert space apparatus, in particular, that each accessible variable has a unique symmetric operator connected to it. These symmetry assumptions will be
shown in this article to be satisfied when all accessible variables take a finite number of values.
\bigskip

One can consider a concrete context with an observer $C$ or with a set of communicating observers in this context. 
Let $\phi$ be an inaccessible theoretical variable varying in a space $\Omega_\phi$. It is a basic philosophy of the present paper that I always regard groups as group actions or transformations, acting on some space.

Starting with $\Omega_\phi$ and a group $K$ acting on $\Omega_\phi$, let $\theta(\cdot)$ be an accessible function on $\Omega_\phi$, and let $\Omega_\theta$ be the range of this function.

As mentioned in the Introduction, I regard `accessible' and `inaccessible' as primitive notions. But they have concrete interpretations, at least in the physical case: Roughly, a physical variable $\theta$ is called accessible if an observer, by a suitable measurement, can obtain as accurate values of it a he wants to. If $\theta$ is accessible, and $\lambda$ is less than or equal to $\theta$ in the partial ordering of variables defined by functional dependence, then $\lambda$ is also accessible.
\bigskip

 \textbf{Definition 1. }
\textit{The accessible variable $\theta$ is called \emph{maximal} if the following holds: $\theta$ is maximal under the partial ordering defined by $\alpha\le \beta$ if $\alpha = f(\beta )$ for some function. In other words: Whenever $\theta = f(\eta )$ for some function $f$ which is not bijective, then $\eta$ is inaccessible.}
\bigskip

\textbf{Postulate 4:} \textit{There exist maximal accessible variables relative to this partial ordering. For every accessible variable $\lambda$ there exists a maximal accessible variable $\theta$ and a function $f$ such that $\lambda = f(\theta)$.}
\bigskip

Two different maximal accessible variables come very close to what Bohr called complementary variables; see Plotnitsky [37] for a thorough discussion.

It is crucial what is meant by `different' here. If $\theta=f(\eta)$ where $f$ is a bijective function, there is a one-to-one correspondence between $\theta$ and $\eta$, they contain the same information, and they must be considered `equal' in this sense. $\theta$ and $\eta$ are said to be `different' if they are not `equal' in this meaning. This is consistent with the partial ordering in Definition 1. The word `different' is used in the same meaning in the Theorems below.

The property of being `equal' in this sense is an equivalence relation. In several applications, just one member of each equivalence class has a physical meaning. In other applications - say if the variable is `time'or `space' - we have the freedom to chooose a unit, so all variables that are linear functions of each other, have equal physical meaning.

Postulate 4 can be proved directly by using Zorn’s lemma - if this lemma, which is equivalent to the axiom of choice, is assumed to hold - and Postulate 3, but such a motivation is not necessary if Postulate 4 is accepted. Physical examples of maximal accessible variables
are the position or the momentum of some particle, or the spin component in some
direction. In a more general situation, the maximal accessible variable may be a vector, whose components are simultaneously measurable.

In example A) the individual spin components can be taken to be maximal. In example B) both position and momentum are maximal as accessible theoretical variables. 

A statistical model for position measurement might be that the measured position is equal to the theoretical position plus noise. In this model, the accessible variable 'theoretical position' can be seen as a statistical parameter.

These postulates are all that I assume. The first goal of this article is to prove through some mathematical arguments versions of the following theorem:
\bigskip

\textbf{Theorem 0: } \textit{Assume that there relative to an observer $C$ in some given context among other variables exist two different maximal accessible variables, each taking $n$ values. Assume that these two are not bijective functions of each other. Then there exists an $n$-dimensional Hilbert space $\mathcal{H}$ describing the situation, and every accessible variable in this situation will have a unique self-adjoint operator in $\mathcal{H}$ associated with it.}
\bigskip

Theorem 0 is identical to Theorem 6 of Section 7, which can be deduced from the more general Theorem 4. Note that the crucial assumption is that we have two - in Niels Bohr's terminology - complementary variables.

These Theorems can be seen as my first starting point for developing the quantum formalism from simple postulates. In one version of Theorem 4, see [4], it is assumed that the two basic variables above are \emph{related}. This can be dispensed with in the finite-dimensional case. In the general case, it will be proved to be a consequence of the assumption that there exists a group $K$ acting upon $\phi$ (see Proposition 1 below).

The property of being related will be defined here:
\bigskip

\textbf{Definition 2.} \textit{Let $\theta$ and $\eta$ be two accessible variables in some context, and let $\theta = f(\phi)$ for some $\phi\in\Omega_\phi$ and some function $f$. If there then is a transformation $k$ of $\Omega_\phi$ such that this holds iff  $\eta(\phi) = f(k\phi)$, we say that $\theta$ and $\eta$ are related relative to this $\phi$. If no such $k$ can be found, we say that $\theta$ and $\eta$ are non-related relative to the variable $\phi$.}
\bigskip

It is easy to show that the property of being related is an equivalence relation. And if $\theta$ is maximal as accessible variable, it follows from the relationship property that $\eta$ above also is maximal. Finally, if $G$ is a group acting on $\Omega_\theta$, and $\theta = f(\phi)$, there can be defined a corresponding group $H$ acting on $\Omega_\eta$ by $h\eta(\phi)=(gf)(k\phi)$ if $\eta(\phi)=f(k\phi)$ and $g\in G$. The mapping from $g$ to $h$ is an isomorphism.

Using the above postulates in the finite-dimensional case, further results can be proved, among other things:

- The eigenvalues of the operator associated with $\theta$ are the possible values of $\theta$.

- The accessible variable $\theta$ is maximal if and only if all eigenvalues are simple.

- The eigenspaces of the operator are associated with one of several variables, say $\theta$. are in one-to-one correspondence with questions of the form ‘What is $\theta$?/ What will $\theta$ be if it is measured?’ together with sharp answers ‘$\theta=u$’ for some $u$. In the maximal case, this gives a simple interpretation of eigenvectors.

To show all this in detail requires separate proofs given in Appendix 2. For the most general version of my approach, I crucially need the assumption (Postulate 3) that there is a group $K$ of actions on the range space $\Omega_\phi$ of $\phi$. Also, for at least one maximal accessible variable $\theta$, I need to assume that there is a group $G$ acting on $\Omega_\theta$ that is transitive and has a trivial isotropy group (see later).

The existence of the group $K$ implies a property of relatedness between two different accessible variable $\theta$ and $\eta$ that have similar ranges.
\bigskip

\textbf{Proposition 1.}
\textit{Assume that the basic inaccessible variable $\phi$ satisfies Postulate 3, and that two given accessible variables $\theta$ and $\eta$ have ranges $\Omega_\theta$ and $\Omega_\eta$ that are in one-to-one correspondence. Then either there is a bijective function connecting $\theta$ and $\eta$, or the following holds: There exists an accessible variable $\xi$ which is a bijective function of $\eta$ such that $\theta$ and $\xi$ are related.}
\bigskip

\textit{Proof.} See Proposition 1 in [82]. $\Box$

In many applications, it turns out that $\xi$ equals $\eta$, so that $\theta$ and $\eta$ are related.
\bigskip

I will also partly need the following:
\bigskip

\textbf{Definition 3.}
\textit{The accessible variable $\theta$ is called\emph{ permissible} with respect to $K$ if the following holds: $\theta(\phi_1)=\theta(\phi_2)$ implies $\theta(t\phi_1 )=\theta(t\phi_2 )$ for all group elements $t\in K$.}
\bigskip

With respect to parameters and subparameters along with their estimation, the concept of permissibility is discussed in some detail in Chapter 3 in [24]. The main conclusion, which also is valid in this setting, is that under the assumption of permissibility, one can define a group $G$ of actions on $\Omega_\theta$ with elements $g$ defined for any $t\in K$ by

\begin{equation}
(g\theta)(\phi):=\theta(t\phi);\ t\in K.
\label{tg}
\end{equation}

Herein I use different notations for the group actions $g$ on $\Omega_\theta$ and the group actions $t$ on $\Omega_\phi$; by contrast, the same symbol $g$ was used in [24]. The background for that is
\bigskip

\textbf{Lemma 1.}
\textit{Assume that $\theta$ is a permissible variable. The function from $K$ to $G$ defined by (\ref{tg}) is then a group homomorphism.}
\bigskip

\textit{Proof.} See [25]. $\Box$
\bigskip

In general, whether the function $\theta(\cdot)$ is permissible or not, I will assume that a transitive group $G$ is acting on $\theta$, and this is enough for the general Theorem 4 below, from which Theorem 6/Theorem 0 can be deduced. To prove the above properties of the Hilbert space formulation, I need a further basic result. Theorem 5 of Section 8 is the general result, and Theorem 7 is a simpler version, valid in the finite-dimensional case.
\bigskip

Note that my approach here can be seen as fully epistemic. It has to do with an agent seeking knowledge. In the finite-dimensional case, we may concentrate on state vectors that are eigenvectors of some meaningful operator. If this operator is associated with a maximal accessible variable $\theta$, then in general these state vectors have interpretations as questions-and-answers: First, look at questions of the form: `What is $\theta$?' or `What will $\theta$ be if we measure it?'. Then consider sharp answers of the form `$\theta=u$', where $u$ is an eigenvalue of the operator corresponding to $\theta$. This is equivalent to specifying an eigenvector of the operator corresponding to $\theta$.

To show this requires some mathematics, given in this article, where also a further discussion is given. What is lacking in this development, are arguments for the Schr\"{o}dinger equation and for the Born formula from simple assumptions. These issues will also be briefly discussed in Section 7 below, and are discussed in more detail in [1].

It is crucial for my development that operators corresponding to the accessible theoretical variables are found first. As a consequence of this article I will consider a version of quantum mechanics where a ket vector is only seen as a state vector when it is an eigenvector of some physically meaningful operator.

The development above was limited to a single observer $C$. Now the same mathematics applies to the following situation: There is a set of communicating observers, and jointly accessible or inaccessible variables in some context are associated with these observers. There may be difficulties, in general, to establish what the basic inaccessible variable  $\phi$ should be in the latter situation, but at least in the two physical examples above the construction is clear. Through discussions the set of observers can establish their theoretical variables, and find out which of them are accessible. The only difference now is that, in order to secure communication, the variables must be possible to define by words.

Note that in this whole discussion, I have said nothing about the ontology. I am fully convinced that there exists an external world, but the detailed properties of this world may be outside our ability to find out. However, in very many cases, the group of observers above may in principle consist of all human beings. Then it seems natural to conclude that the physical phenomenon studied also can be seen as an ontological phenomenon.

Quantum mechanics as a model, although it is a very good model, can sometimes only give partial answers. Ontological aspects of my approach are further discussed in [73].

I admit that this approach is unusual and that the postulates to some may seem a little unusual. However, for an outsider, I will claim that it is much easier to understand these postulates than jumping right into the usual Hilbert space formalism. For those of us who have learned formalism, the approach may in some sense require some unlearning first.

Possible relationships between my approach and some other approaches towards the foundation of quantum mechanics will be briefly discussed in Section 8 below. My assumptions seem to be weaker than many other basic assumptions made.

The theory of this paper can be extended to relativistic quantum field theory; see [84]. On slightly extending the postulates, the theory also seems to have consequences for psychology, sociology, and theology; see [83]. This is consistent with Andrei Khrennikov's assumption of an ubiquitous quantum structure [89]. The theory and some of its consequences are also discussed in [90].

\section{Group actions and measures}

Starting with a point $\theta_0\in \Omega_\theta$, an orbit of a group $G$ acting on $\Omega_\theta$ is the set $\{g\theta_0:\ g\in G\}$. It is trivial to see that the orbits are disjoint, and their union is the full space $\Omega_\theta$. The point $\theta_0$ may be replaced by any point of the same orbit. In the case of one orbit filling the whole space, the group is said to be transitive.

The isotropy group at a point $\theta\in\Omega_\theta$ is the set of $g$ such that $g\theta=\theta$. It is easy to see that this is a group. If this group is trivial for one $\theta = \theta_1$, it is trivial for all $\theta$ in the orbit containing $\theta_1$.

It is important to define left and right invariant measures, both on the groups and on the spaces of theoretical variables. In the mathematical literature, see for instance [26,27], Haar measures on the groups are defined (assuming locally compact groups). Right ($\mu_G$) and left ($\nu_G$) Haar measures on the group $G$ satisfy
\begin{eqnarray*}
\mu_G(Dg)=\mu_G(D), \ \mathrm{and}\ \nu_G(gD)=\nu_K(D)\\ \mathrm{for}\ g\in G\ \mathrm{and}\ D\subset G,\ \mathrm{respectively}.
\end{eqnarray*}

Next, define the corresponding measures on $\Omega_\theta$. As is commonly done, I assume that the group operations $(g_1,g_2)\mapsto g_1g_2$, $(g_1,g_2)\mapsto g_2g_1$ and $g\mapsto g^{-1}$ are continuous. Furthermore, I will assume that the action $(g,\theta)\mapsto g\theta$ is continuous. 

As discussed in Wijsman [28], an additional condition is that every inverse image of compact sets under the function $(g,\theta)\mapsto (g\theta,\theta)$ should be compact. A continuous action by a group $G$ on a space $\Omega_\theta$ satisfying this condition is called \emph{proper}. This technical condition turns out to have useful properties and is assumed throughout this paper. When the group action is proper, the orbits of the group can be proved to be closed sets relative to the topology of $\Omega_\theta$.

The following result, originally due to Weil, is proved in [26,28]; for more details on the right-invariant case, see also [24].
\bigskip

\textbf{Theorem 1.}
\textit{The left-invariant measure $\nu$ on $\Omega_\theta$ exists if the action of $G$ on $\Omega_\theta$ is proper and the group is locally compact.}
\bigskip

The connection between $\nu_G$ defined on $G$ and the corresponding left invariant measure $\nu$ defined on $\Omega_\theta$ is relatively simple: If for some fixed value $\theta_0$ of the theoretical variable the function $\beta$ on $G$ is defined by $\beta: g\mapsto g\theta_0$, then $\nu(E)=\nu_G (\beta^{-1}(E))$.This connection between $\nu_G$ and $\nu$ can also be written $\nu_G(dg)=d\nu(g\theta_0))$, so that $d\nu(hg\phi_0)=d\nu (g\phi_0)$ for all $h, g \in G$ if $\nu$ is left-invariant..

Note that $\nu$ can be seen as an induced measure on each orbit of $G$ on $\Omega_\theta$, and it can be arbitrarily normalized on each orbit. $\nu$ is finite on a given orbit if and only if the orbit is compact. In particular, $\nu$ can be defined as a probability measure on $\Omega_\theta$ if and only if all orbits of $\Omega_\theta$ are compact. Furthermore, $\nu$ is unique only if the group action is transitive. Transitivity of $G$ as acting on $\Omega_\theta$ will be assumed throughout this paper.

In a corresponding fashion, a right invariant measure can be defined on $\Omega_\theta$. This measure satisfies $d\mu (gh\phi_0)=d\mu (g\phi_0)$ for all $g,h\in G$. In many cases the left invariant measure and the right invariant measure are equal.

\section{A brief discussion of group representation theory}
 
A group representation of $G$ is a continuous homomorphism from $G$ to the group of invertible linear operators $V$ on some vector space $\mathcal{H}$:
\begin{equation}
V(g_1 g_2 )=V(g_1 )V(g_2 ).
\label{3}
\end{equation}
It is also required that $V(e)=I$, where $I$ is the identity, and $e$ is the unit element of $G$. This assures that the inverse exists: $V(g)^{-1}=V(g^{-1})$. The representation is unitary if the 
operators are unitary ($V(g)^{\dagger}V(g)=I$). If the vector space is finite-dimensional, we have a representation $D(V)$ on the square, invertible matrices. For any 
representation $V$ and any fixed invertible operator $U$ on the vector space, we can define a new equivalent representation as $W(g)=UV(g)U^{-1}$. One can prove that two 
equivalent unitary representations are unitarily equivalent; thus $U$ can be chosen as a unitary operator.

A subspace $\mathcal{H}_1$ of $\mathcal{H}$ is called invariant with respect to the representation $V$ if $u\in \mathcal{H}_1$ implies $V(g)u\in \mathcal{H}_1$ for all $g\in G$. The null-space $\{0\}$ and the whole space
$\mathcal{H}$ are trivially invariant; other invariant subspaces are called proper. A group representation $V$ of a group $G$ in $\mathcal{H}$ is called irreducible if it has no proper invariant subspace.
A representation is said to be fully reducible if it can be expressed as a direct sum of irreducible subrepresentations. A finite-dimensional unitary representation of any group 
is fully reducible. In terms of a matrix representation, this means that we can always find a  $W(g)=UV(g)U^{-1}$ such that $D(W)$ is of minimal block diagonal form. Each one of 
these blocks represents an irreducible representation, and they are all one-dimensional if and only if $G$ is Abelian. The blocks may be seen as operators on subspaces of the 
original vector space, i.e., the irreducible subspaces. The blocks are important in studying the structure of the group.

A useful result is Schur's Lemma; see for instance [27]:
\bigskip

\textit{Let $V_1$ and $V_2$ be two irreducible representations of a group $G$; $V_1$ on the space $\mathcal{H}_1$ and $V_2$ on the space $\mathcal{H}_2$. Suppose that there exists a linear map $T$ from $\mathcal{H}_1$ to 
$\mathcal{H}_2$ such that}
\begin{equation}
V_2 (g)T(v)=T(V_1 (g)v)
\label{4}
\end{equation}
\textit{for all $g\in G$ and $v\in\mathcal{H}_1$.}

\textit{Then either $T$ is zero or it is a linear isomorphism. Furthermore, if $\mathcal{H}_1=\mathcal{H}_2$ and $V_1 = V_2$, then $T=\lambda I$ for some complex number $\lambda$.}
\bigskip

Let $\nu$ be the left-invariant measure of the space $\Omega_\theta$ induced by the group $G$, and consider in this connection the Hilbert space $\mathcal{H}=L^2 (\Omega_\theta ,\nu)$. Then the left-regular 
representation of $G$ on $\mathcal{H}$ is defined by $U^{L}(g)f(\theta)=f(g^{-1}\theta)$. This representation always exists, and it can be shown to be unitary, see [29]. I assume that $G$ is transitive on $\Omega_\theta$ and has a trivial isotropy group.
\bigskip

\textbf{Proposition 2.}
\textit{There exists a $|\theta_0\rangle \in L^2 (\Omega_\theta ,\nu)$ such that the coherent states $U^L (g)|\theta_0\rangle$ for $g\in G$ are in one-to-one correspondence with $\theta\in \Omega_\theta$.}
\bigskip

\textit{Proof.} See Proposition 2 in [82]. $\Box$

For references to some of the vast literature on group representation theory, see Appendix A.2.4 in [24].

\section{The construction of operators for the hypothetical case of an irreducible representation of the basic group}

In the quantum-mechanical context defined in [1] and discussed above, $\theta$ is an accessible variable, and one should be able to introduce an operator associated with $\theta$. The following discussion, which is partly inspired by [29, 31], assumes first an irreducible unitary representation of $G$ on a complex Hilbert space $\mathcal{H}$. In the next Section, the assumption of irreducibility will be removed, by simply assuming that we have two related maximal accessible variables in the given context.

\subsection{A resolution of the identity}

In the following, I assume that the group $G$ has representations that give square-integrable coherent state systems (see page 43 of [29]). For instance, this is the case for all representations of compact semisimple groups, representations of discrete series for real semisimple groups, and some representations of solvable Lie groups.

Let $G$ be an arbitrary such group, and let $V(\cdot)$ be one of its unitary irreducible representations acting on a Hilbert space $\mathcal{H}$. Assume that $G$ is acting transitively on the space $\Omega_\theta$, and fix $\theta_0\in\Omega_\theta$. Then every $\theta\in\Omega_\theta$ can be written as $\theta=g\theta_0 $ for some $g\in G$. I also assume that the isotropy groups of $G$ are trivial. Then this establishes a one-to-one correspondence between $G$ and $\Omega_\theta$. In particular, this implies that the group action is proper and a left-invariant measure $\nu$ on $\Omega_\theta$ exists; see Theorem 1 above.

Also, fix a vector $|\theta_0\rangle\in\mathcal{H}$, and define the coherent states $|\theta\rangle=|\theta(g)\rangle=V(g)|\theta_0\rangle$. With $\nu$ being the left invariant measure on $\Omega_\theta$, introduce the operator
\begin{equation}
T=\int |\theta(g)\rangle\langle\theta(g)|d\nu(g\theta_0).
\label{5}
\end{equation}
Note that the measure here is over $\Omega_\theta$, but the elements are parametrized by $G$. $T$ is assumed to be a finite operator.
\bigskip

\textbf{Lemma 2.}
\textit{$T$ commutes with every $V(h); h\in G$.} 
\bigskip

\textit{Proof.} $\ \ V(h)T=$
\begin{eqnarray*}
\int V(h) |\theta(g)\rangle\langle\theta(g)|d\nu(g\theta_0)
=\int |\theta(hg)\rangle\langle\theta(g)|d\nu(g\theta_0)\\=\int |\theta(r)\rangle\langle\theta(h^{-1}r)|d\nu(h^{-1}r\theta_0 ).
\end{eqnarray*}
Since $|\theta(h^{-1}r)\rangle=V(h^{-1}r)|\theta_0\rangle =V(h^{-1})V(r)|\theta_0\rangle =V(h)^\dagger |\theta(r)\rangle$, we have $\langle \theta(h^{-1}r)|=\langle\theta(r)|V(h)$, and since the measure $\nu$ is left-invariant, it follows that $V(h)T=TV(h)$. $\Box$
\bigskip

From the above and Schur's Lemma, it follows that $T=\lambda I$ for some $\lambda$. Since $T$ by construction only can have positive eigenvalues, we must have $\lambda >0$. Defining the measure $d\rho(\theta)=\lambda^{-1}d\nu(\theta)$ we therefore have the important resolution of the identity
\begin{equation}
\int |\theta\rangle\langle\theta |d\rho(\theta)=I.
\label{6}
\end{equation}
For a more elaborate similar construction taking into account the isotropy subgroups, see Chapter 2 of [31]. In [4] a corresponding resolution of the identity is derived for states defined through representations of the group $K$ acting on $\Omega_\phi$. 

\subsection{Simple quantum operators}

Let now $\theta$ be a maximal accessible variable and let $G$ be a group acting on $\theta$, satisfying the requirements of the last subsection.

In general, an operator corresponding to $\theta$ may be defined by
\begin{eqnarray}
A^\theta
=\int \theta |\theta\rangle\langle\theta | d\rho (\theta).
\label{7}
\end{eqnarray}
$A^\theta$ is defined on a domain $D(A^\theta)$ of vectors $|v\rangle\in\mathcal{H}$ where the integral defining $\langle v|A^\theta|v\rangle$ converges.

This mapping from an accessible variable $\theta$ to an operator $A$ has the following properties:

(i) If $\theta=1$, then $A^\theta=I$.

(ii) If $\theta$ is real-valued, then $A^\theta$ is symmetric (for a definition of this concept for operators and its relationship to self-adjointness, see Appendix 2.)

(iii) The change of basis through a unitary transformation is straightforward.

For further important properties, we need some more theory. First, consider the situation where we regard the group $G$ as generated by a group $K$ defined on the space of an inaccessible variable $\phi$. This represents no problem if the mapping from $\phi$ to $\theta$ is permissible, a case discussed in [4], and in this case, the operators corresponding to several accessible variables can be defined on the same Hilbert space. In the opposite case, we have the following theorem.
\bigskip

\textbf{Theorem 2.} \textit{Let $H$ be the subgroup of $K$ consisting of any transformation $h$ such that $\theta(h\phi)=g\theta(\phi)$ for some $g\in G$. Then $H$ is the maximal group under which the variable $\theta$ is permissible.}
\bigskip

\textit{Proof.} Let $\theta(\phi_1) =\theta(\phi_2)$ for all $\theta\in\Theta$. Then for $h\in H$ we have $\theta(h\phi_1)=g\theta(\phi_1)=g\theta(\phi_2)=\theta (h\phi_2)$, thus $\theta$ is permissible under the group $H$. For a larger group, this argument does not hold. $\Box$
\bigskip

Next look at the mapping from $\theta$ to $A = A^\theta$ defined by (\ref{7}). 
\bigskip

\textbf{Theorem 3.} \textit{ For $g\in G$, $V(g^{-1})AV(g)$ is mapped by $\theta'=g\theta$.}
\bigskip

\textit{Proof.} $V(g^{-1})AV(g)=$
\begin{eqnarray*}
\int \theta |g^{-1}\theta\rangle\langle g^{-1}\theta |d\rho(\theta)=\int g\theta |\theta\rangle\langle \theta |d\rho(g\theta).
\end{eqnarray*}
Use the left invariance of $\rho$. $\Box$
\bigskip

Further properties of the mapping from $\theta$ to $A^\theta$ may be developed in a similar way.
The mapping corresponds to the usual way that the operators are allocated to observables in the quantum mechanical literature. But note that this mapping comes naturally here from the notions of theoretical variables and accessible variables on which group actions are defined.

\section{The main theorems and theoretical results.}

\subsection{The general case}

Up to now, I have assumed an irreducible representation of the group $G$. A severe problem with this, however, is that the group $G$ in many applications is Abelian, and Abelian groups have only one-dimensional irreducible representations. Then the above theory is trivial.

In [4] this problem is solved by taking as a point of departure \emph{two different related maximal accessible variables} $\theta$ and $\eta$. Recall here the meaning of the word `different' discussed in Section 3. The main result is then as follows.
\bigskip

\textbf{Theorem 4} \textit{Consider a context where there are two different maximal accessible variables $\theta$ and $\eta$. Assume that both $\theta$ and $\eta$ are real-valued or real vectors, taking at least two values.  Assume Postulates 1-4, and make the following additional assumptions:}
\smallskip

\textit{(i) On one of these variables, $\theta$, there can be defined a transitive group of actions $G$ with a trivial isotropy group and with a left-invariant measure $\rho$ on the space $\Omega_\theta$.}

\textit{(ii) The ranges $\Omega_\theta$ and $\Omega_\eta$ are in one-to-one correspondence.}

\textit{Then there exists a Hilbert space $\mathcal{H}$ connected to the context, and on $\mathcal{H}$ there exist symmetric operators $A^\theta$ and $A^\eta$ defined by (\ref{x1}) and (\ref{x2}) below.}

 \textit{ If these operators are self-adjoint for every choice of $\eta$, then to every accessible variable in the context, there is associated a self-adjoint operator on $\mathcal{H}$.}
\bigskip

For precise definitions of the concepts `symmetric operator' and `self-adjoint operator', and for conditions under which a symmetric operator is self-adjoint, see [32] and Appendix 2. 

The crucial point in the proof of the first part of Theorem 4 is to construct a group $N$ acting on the vector $\psi =(\theta, \eta)$, and then a representation $W(\cdot)$ of $N$ which I prove is irreducible. The coherent states $|v_n\rangle=W(n)|v_0\rangle$ are then in one-to-one correspondence with $n\in N$. For the details of all this, I refer to Appendix 1 and to [4]. 

This gives the crucial identity
\begin{equation}
\int |v_n\rangle\langle v_n |\nu (dn) = I,
\label{x0}
\end{equation}
where $\nu$ is a left-invariant measure on the group $N$.

One can show that there is a function $f_\theta$ on $N$ such that $\theta=f_\theta (n)$, and a function $f_\eta$ on $N$ such that $\eta=f_\eta (n)$. We can now define operators corresponding to $\theta$ and $\eta$:

\begin{equation}
A^\theta=\int f_\theta (n)|v_n\rangle\langle v_n |\nu (dn),
\label{x1}
\end{equation}

\begin{equation}
A^\eta=\int f_\eta (n)|v_n\rangle\langle v_n |\nu (dn).
\label{x2}
\end{equation}

Note that any pair of maximal accessible variables may be used as a basis for Theorem 4.
Accessible variables that are not maximal, can always be seen as functions of a maximal variable by Postulate 4. Hence for these variables, the spectral theorem may be used, and operators constructed as in Appendix 2.

An essential part of the proof of Theorem 4 is to prove that if a certain representation $U(\cdot)$ of $G$  is not irreducible, then $W(\cdot )$ is an irreducible representation of $N$. In order to carry out this part of the proof, I need a representation $U(\cdot )$ which is at least three-dimensional, so that it can be reduced to a lower-dimensional space if not irreducible, and similarly the representation of the corresponding group $H$ acting upon $\eta$ must be at least three-dimensional. (The two-dimensional case, the qubit case, is treated separately in [1].)

Theorem 4 is first proved under the assumption that $\theta$ and $\eta$ are related as i Definition 2. But by Proposition 1, this represents no limitation if we are allowed to replace $\eta$ by a bijective function of $\eta$. To show that this may be permitted, we can use the spectral theorem again; see Appendix 2.

So assume that the two variables are related. The transformation $k$ defining $\eta=f(k\phi)$ from $\theta=f(\phi)$ cannot be just the trivial one interchanging $\theta$ and $\eta$, taken together with the assumption that the group $G$ acting upon $\theta$ is just the identity. This is clear, since if such a trivial interchange was allowed in this case, every pair of variables would be related by the above definitions. Note, however, that according to Definition 2, the notion of being related depends on the inaccessible variable $\phi$. By Postulate 1 there is a group $K$ defined on the range of this variable.

If we for instance take $\phi$ as the vector (position, momentum), and have a sufficiently large group $G$ connected to $\theta$=position (say, the translation group), then an interchange of position and momentum is permitted \emph{relative to this $\phi$.} More concretely, this interchange will involve a Fourier transform, as in the ordinary theory. The group $K$ acting on $\phi$ in this case may be taken as the Heisenberg-Weyl group; see [31].

To complete the construction of the usual Hilbert space formalism from the mathematical model of Section 2, I need a further main theorem.
\bigskip

\textbf{Theorem 5}
\textit{(i) If $\theta$ and $\eta$ are different, but related through a transformation $k$ of $\Omega_\phi$, there is a unitary operator $S(k)$ such that $A^\eta = S(k)^\dagger A^\theta S(k)$.}

\textit{(ii) Assume that the functions $\theta(\cdot)$ and $\eta(\cdot)$ are permissible with respect to a group $K$ acting on $\Omega_\phi$. Assume that $K$ is transitive and has a trivial isotropy group. Let $T(\cdot)$ be a unitary representation of $K$ such that the coherent states $T(t)|\psi_0\rangle$ are in one-to-one correspondence with $t$. For any transformation $t\in K$ and any such unitary representation $T$ of $K$, the operator $T(t)^\dagger A^\theta T(t)$ is the operator corresponding to $\theta'$ defined by $\theta'(\phi)=\theta(t\phi)$.}

\bigskip

This is also proved in Appendix 1.

One final remark to the developments above: The above theorems have so far been connected to a single observer $A$ and the mathematical model of Section 3. But the same arguments can be used with the following model: Assume a set of communicating observers and assume that these have defined joint variables that may be accessible or inaccessible to the set of observers. Then the same mathematics is valid, and the same mathematical/physical examples of variables may be used.

  \begin{figure}[t]
\includegraphics{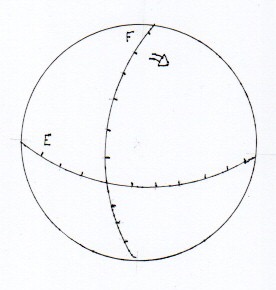}
{\caption{The construction of the transformation k.\label{fig1}}}
 \end{figure}

\subsection{The case where the maximal accessible variable takes a finite number of values}

I will show here first that if $\theta$ and $\eta$ take a finite $n$ number of values, then we can choose $G$ such that the symmetry assumptions of Theorem 4 are satisfied, and we can always choose $k$ such that the two variables are related. This leads to a simplification of the theory. The case $n=2$, the qubit case, is discussed separately in [1]; see Subsection 4.5.3 and Section 5.2 there. I will here assume $n\ge 3$, which in fact can be shown to be needed in the proof in Appendix 1 of Theorem 4.

In the finite case, it is crucial that the reducibility of the representation $U(\cdot )$ is permitted.
Concretely, let $G$ be the cyclic group acting on the distinct values $u_1,...,u_n$ of $\theta$, that is, the group generated by the element $g_0$ such that $g_0 u_i =u_{i+1}$ for $i=1,..., n-1$ and $g_0 u_n = u_1$. This is an Abelian group, which only has one-dimensional irreducible representations. However, we can define $U(\cdot )$ as taking values as diagonal unitary $n\times n$ matrices with different complex $n$th roots of the identity on the diagonal. For the specific matrix $U(g_0)$, take these $n$th roots in their natural order, and then let every element of $G$ be mapped into the diagonal matrices $U(\cdot )$ by the corresponding cyclical permutation. 

It is easy to see then that the coherent states $U(g)|\theta_0\rangle$  are in one-to-one correspondence with the group elements $g\in G$ when  $|\theta_0\rangle$ is a unit vector with one element equal to 1 and the others zero, and this can be generalized to any $|\theta_0\rangle$. Also, $G$ is transitive in its range and has a trivial isotropy group. 

I will also prove here that $\eta$ can be found as a related variable to $\theta$, that is, the existence of an inaccessible variable $\phi$ and a transformation $k$ in the corresponding space $\Omega_\phi$ such that $\eta(\phi)=\theta(k\phi)$.

To this end, let $\Omega_\phi$ be the three-dimensional unit sphere, plot the values of $\eta$ along the equator $E$, and the values of $\theta$ along the great circle $F$ containing the south pole and the north pole. See Figure 1. 

  \begin{figure}[t]
\includegraphics{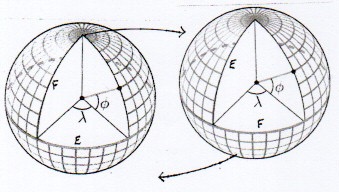}
{\caption{The construction of the group K, acting on grids on two copies of the sphere.\label{fig2}}}
 \end{figure}

Without loss of generality, we can let the values of $\theta$ and $\eta$ be equidistant. (If this is not the case, we can define equivalent variables $\theta$ and $\eta$ that are equidistant. If operators are proved to exist for these equivalent variables, operators of the original variables can be constructed from the spectral theorem.The spectral theorem is trivially valid in the finite-dimensional case.) If these values are plotted in a corresponding way, we can transform the values of $\theta$ onto the values of $\eta$ by a $90^o$ rotation $k$ of the sphere as indicated in the figure. Thus $\eta(\phi)=\theta(k\phi)$. All the symmetry assumptions of Theorem 4 are satisfied, and we have simply
\bigskip

\textbf{Theorem 6} \textit{Assume Postulate 1 to Postulate 4 of Section 3, and that there exist two different maximal accessible variables $\theta$ and $\eta$, each taking $n\ge 3$ values, and not being bijective functions of each other. Then, there exists an $n$-dimensional Hilbert space $\mathcal{H}$ describing the situation, and every accessible variable in this situation will have an associated self-adjoint operator in $\mathcal{H}$.}
\bigskip

This is a very simple and strong result, which can be taken as a basis for a large part of textbook quantum mechanics. Note that in the finite-dimensional case, all operators are bounded, and the notions symmetric and self-adjoint operators coincide.

According to Postulate 3 of Section 3, there always exists a group $K$ acting on $\Omega_\phi$. One question is now whether this group can be constructed such that the symmetry assumptions of Theorem 5(ii) are satisfied. I will only indicate a constrution here, since Theorem 5(ii) is not used in the final theory. Take as a basis the cyclic group $G$ acting on the values of $\theta$, and let a corresponding group $H$ act on the values of $\eta$. Without loss of generality, assume these values to be equidistant. 

Construct $K$, acting on two copies of the sphere as in Figure 2 as follows: Make a grid on both spheres by discretizing the longitude angle $\lambda$ and the latitude angle $\phi$ as on the Figure (not to be confused with the basic inaccessible variable $\phi$). In the first copy, this includes the lines $E$ (representing $\theta$) and $F$ (representing $\eta$). In the second copy, $E$ ($\theta$) and $F$ ($\eta$) are switched. The states are the intersections of meridians and latitude circles. In the first copy, let $G$ act on the latitude circles and $H$ act on the meridians. In the second copy, $G$ and $H$ are switched. This gives a version of $G\otimes H$ acting on both copies. We let $K$ consist of this and the following elements: From the North pole and the South pole of both copies there is a group action element $j$ going from one copy to an arbitrarily chosen state of the other. This state can be chosen by a uniform distribution.
     
The representation $U(\cdot )$ of $G$ is taken as above, we construct a similar representation $V(\cdot )$ of $H$, and we take $T(K)=W(K)$ be the irreducible group used in the proof of Theorem 4, where $K$ is extended as above.

Using this geometry, the following must be proved, but proofs are omitted here:

1) $K$ is transitive on its values and has a trivial isotropy group.

2) $\theta(\cdot)$ and $\eta(\cdot)$ are permissible functions of the state variable $\phi$ with respect to $K$.

3) Taking as $|\psi_0\rangle$ one of the points where the great circles $E$ and $F$ intersect, the coherent states $T(t)|\psi_0\rangle$ are in one-to-one correspondence with the group elements $t\in K$.

4) $T(\cdot )$ is unitary and irreducible.

From this, the conditions of the second part of Theorem 5 are satisfied, and we have:
\bigskip

\textbf{Theorem 7}
\textit{(i) For the special transformation $k$ above, we have $A^\eta =S(k)^\dagger A^\theta S(k)$ for some unitary matrix $S(k)$. .}

\textit{(ii) Assume Postulate 1 to Postulate 4 of Section 3, and that there exist two different accessible variables $\theta$ and $\eta$, each taking $n$ values. Let $T(\cdot)$ be the unitary representation of $K$ defined above. Then for any transformation $t\in K$, the operator $T(t)^\dagger A^\theta T(t)$ is the operator corresponding to $\theta'$ defined by $\theta'(\phi)=\theta(t\phi)$.}
\bigskip

The first part of Theorem 7 is just a repetition of the first part of Theorem 5.

From these two theorems follow a rich class of results, as discussed in detail in [1]; see the proofs of two theorems in Appendix 3. (The second part of Theorem 7 is not needed for these results):

- Every accessible variable has a self-adjoint operator connected to it. 

- The set of eigenvalues of the operator is equal to the set of possible values of the variable.

- An accessible variable is maximal if and only if all eigenvalues are simple.

- The eigenvectors can, in the maximal case, be interpreted in terms of a question together with its answer. Specifically, this means that in a context with several variables, a chosen maximal variable $\theta$ may be identified with the question `What will $\theta$ be if we measure it?' and a specific eigenvector of $A^\theta$, corresponding to the eigenvalue $u$ may be identified with the answer `$\theta=u$'.

- In the general case, eigenspaces have the same interpretation.

- The operators of related variables are connected by a unitary similarity transform.
\bigskip

For the proofs of the second and third statements above, see Appendix 3. The third and fourth statements can be taken as basis for my proposal of a version of quantum mechanics: A ket vector is only seen as a valid state vector if it is an eigenvector of a physically meaningful operator. This requires a separate discussion of the superposition principle,

In my theory, the only valid state vectors are related to some accessible variable $\eta$ and the possible values $v_k$ of this $\eta$. Assume for simplicity that $\eta$ is maximal. Then the statement $\eta=v_k$ corresponds to an eigenvector $|v_k\rangle$ of $A^\eta$, with the resolution of the identity
\begin{equation}
\sum_k |v_k\rangle\langle v_k |=I.
\label{ridentity}
\end{equation}

This implies a large class of possible new state vectors
\begin{equation}
| \psi\rangle = \sum_k |v_k\rangle\langle v_k | \psi \rangle = \sum_k \langle v_k | \psi\rangle| v_k \rangle =\sum_k  c_k |v_k\rangle , 
\label{superp}
\end{equation}
but it does not follow from this that every superposition of the orthogonal ket vectors $|v_k\rangle$ can be written in this way.

Note that $|\psi\rangle$ may also have an interpretation in terms of another accessible variable $\theta$ and some statement $\theta=u$. The physical situation may be that we know something about $\theta$ or know something about $\eta$, but we can also be without such knowledge. The most definite statement about knowledge of $\eta$ is of the form $\eta=v$, but we may also know probability statements of the form $\{ \pi (v_k) \}$, which is formalized by a density matrix
\begin{equation}
\rho = \sum_k \pi(v_k) |v_k\rangle\langle v_k |
\label{densitym}
\end{equation}
As is well known, this leads by Born's formula to probability statements for $\theta$: 
\begin{equation}
P[\theta=u]=\langle \psi | \rho |\psi \rangle =\sum_k \pi(v_k) |\langle \psi |v_k\rangle |^2.
\label{probab}
\end{equation}

\bigskip

Going back to the superposition principle: Let $| \alpha\rangle$ and $| \beta\rangle$ be two different state vectors. Then $| \alpha\rangle$ can be connected to a statement $\gamma=u$ for some theoretical variable $\gamma$, or equivalently, $c(\gamma)=c(u)$ for any bijective function $c$, and  $| \beta\rangle$ can be connected to a statement $\xi=v$ for some theoretical variable $\xi$, or equivalently, $d(\xi)=d(v)$ for any bijective function $d$. By Postulate 4, if $\gamma$ and $\xi$ are not maximal, they are functions, say $f_1$ and $f_2$, of some maximal variables, say $\theta$ and $\eta$, respectively. 
\begin{equation}
\gamma = f_1(\theta),\ \ \xi =f_2 (\eta).
\label{gamma1}
\end{equation}

We have two possibilities. Either $\theta$ and $\eta$ are bijective functions of each other. Then every states connected to them can be expressed in the same basis $\{ | v_k\rangle \}$. Or they are different in the sense defined in Section 3. Assume that the weak conditions of Theorem 4 hold. Then by this theorem  they can be taken to construct the Hilbert space and the necessary operators.

In either case we have by (\ref{x1}) and (\ref{x2}) and the spectral theorem,
\begin{equation}
a A^\gamma + b A^\xi = \sum_k (af_1(u_k)+bf_2(v_k)) |v_k\rangle\langle v_k |,
\label{Aop}
\end{equation}
This is the operator of the theoretical variable $\lambda=a\gamma+b\xi$, or more generally $\lambda=ac(\gamma)+bd(\xi)$ for some bijective mappings $c$ and $d$, and every state $|w\rangle$ associated with $\lambda=w$ can be expressed as
\begin{equation}
|w\rangle =\sum_k r_k |v_k\rangle ,
\label{wstate}
\end{equation}
\smallskip

\textbf{Postulate 5:} \textit{ If $\{u_i \}; i\in I$ are the possible values of the accessible variable $\gamma$, $\{ v_j\}; j\in J$ are the possible values of the accessible variable $\xi$, and $\lambda = f(\gamma,\xi)$ is accessible, then the possible values of $\lambda$ is contained in $\{f(u_i, v_j\}; i\in I, j\in J$.}
\bigskip

\textbf{Theorem 8} \textit{Assume that $|\alpha\rangle$ and $|\beta\rangle$ are possible state vectors, and $|\alpha\rangle$ can be associated with an event $f_1(\theta)=u$, and $|\beta\rangle$ can be associated with an event $f_2(\eta)=v$. Here $\theta$ and $\eta$ are two different  meaningful maximal accessible variables. Then $a|\alpha\rangle+b|\beta\rangle$, $(a\ne 0, b\ne 0)$ is a possible state vector if and only if there exist bijective functions $c$ and $d$ such that $\lambda = ac(f_1(\theta))+bd(f_2(\eta))$ is an accessible variable. .}
\bigskip

\underline{Proof.} As above, if $|\alpha\rangle$ is interpreted as $\gamma=u$, and $|\beta\rangle$ is interpreted as $\xi =v$, this implies an interpretation of $a|\alpha\rangle+b|\beta\rangle$ as $\lambda = ac(\gamma) + bd(\xi )= w = au+bv$ for some bijective mappings $c$ and $d$, and this $\lambda$ is of the desired form. It is left to prove the `only if' part. To this end, assume that $\theta$. $\eta$ and $\lambda$ have the assumed properties. Without loss of generality, let $c$ and $d$ be identities. By Postulate 4 there exist a maximal variable $\mu$ such that $\lambda$ is a function of $\mu$. Since $\theta$ and $\eta$ are different, and $\mu$ must involve $\eta$ in a non-trivial way, $\theta$ and $\mu$ must be different. Then by Theorem 6 we can construct a Hilbert space based upon $\theta$ and $\mu$, operators $A^\mu$ and then the operator $A^\lambda$ by the spectral theorem  correponding to the meaningful variable $\lambda$ exist. Let $\{w_k\}$ be the possible values of $\lambda$.

Associated with $\lambda=w_k$ there must according to Postulate 5 be an $i=i(k)$ and a $j=j(k)$ such that $w_k =au_{i(k)}+bv_{j(k)}$. Here $u_{i(k)}$ is a possible value of $\gamma$, an eigenvalue of $A^\gamma$ with a corresponding eigenvector $|\alpha_{i(k)}\rangle$, and  $v_{j(k)}$ is a possible value of $\xi$, an eigenvalue of $A^\xi$ with a corresponding eigenvector $|\beta_{j(k)}\rangle$. This gives that each $a|\alpha_{i(k)}\rangle+b|\beta_{j(k]}\rangle$ is a possible state vector, corresponding to $\lambda=w_k$.
$\Box$
\bigskip

Note the assumption in Theorem 8. For instance, in the Schr\"{o}dinger cat example. let $|\alpha_1\rangle$ be the state corresponding to a cat which is known to be dead, and let $|\beta_1\rangle$ corresponding to a cat which is known to be alive on suitable Hilbert spaces. Let $|\alpha_2\rangle$ and $|\beta_2\rangle$ correspond to the opposite of these two events, i.e., $|\alpha_2\rangle$ has the interpretation $\gamma_2 = 1$, where $\gamma_2$ is the indicator that the cat is not knwn to be dead. Denote in general the indicators of the mentioned events $\gamma_i$ and $\xi_j$. for $i,j = 1,2$

An observer outside the box knows nothing, and is associated with the state $||\alpha_2\rangle\otimes|\beta_2\rangle$, corresponding to $\gamma_2 = \xi_2 =1$. Any superposition of the states  $|\alpha_i\rangle$ and $|\beta_j\rangle$ is for him meaningless. An observer inside the box, wearing a gas mask, will know the answers, and is associated with the state given by $\gamma_1=1$ or $\xi_1=1$ Again, superposition is meaningless. The two observers will agree on the status of the cat once the door to the box is opened.
\bigskip

It is crucial now that this full theory seems to follow by - in addition to the simple Postulate1 to Postulate 4 of Section 3 - essentially only assuming that two different maximal accessible variables exist, in Niels Bohr's terminology, the existence of two different complementary variables. Born's formula requires additional postulates, as briefly discussed in Subsection 7.6, and more fully in [1,88].

\subsection {The case of position and momentum of a particle}

Theorem 4 and Theorem 5 are valid also in the case that $\theta$ and $\eta$ are continuous theoretical variables..
It is of interest also to develop further the basis of quantum theory for this case, but this is outside the scope of the present paper. See Hall [32] for a completely rigorous theory. But it is fairly straightforward now to complete the theory for an important special case: Let $\theta$ be the theoretical position of some particle and let $\eta$ be its theoretical momentum. I choose the accessible variables to be such theoretical variables and assume that a measurement consists of a theoretical value plus a measurement error. This is similar to how measurements are modeled in statistics. 

The simplest approach is the following: Approximate $\theta$ with an $n$-valued variable $\theta_n$, find an operator $A_n$ corresponding to $\theta_n$, and let $n$ tend to infinity. This approach is carried out in Section 5.3 in [1]. It is shown that the Hilbert space for $\theta$ can be taken to be $L^2(\mathbb{R}, dx)$, and the transformation $k$, which gives the operator for momentum, is a Fourier transform on this Hilbert space. The operators connected to $\theta$ and $\eta$ are the usual ones.
 
A more direct approach, using the general theory here, is to take the group $G$ acting on $\theta$ to be the translation group, and let the group $K$ acting on $\phi =(\theta,\eta)$ be the Heisenberg-Weyl group; see [31]. This will not be further discussed here.

\subsection{Quantum decision theory}

There is a large literature on quantum decision theory; see for instance the survey article [22], the book [23] and the series of articles [38-42] by Yukalov and Sornette. The whole field of quantum decisions can be linked to the theory introduced here, as discussed in [5]. The clue is to let my variables $\theta,\eta,\xi,...$ no longer be physical variables, but decision variables. In the simplest case, a decision variable takes a finite number of values.

Let a person $A$ be in a concrete decision situation. He is among other things faced with the choice between taking actions $a_1,...,a_n$. Define a decision variable $\theta$ as equal to $j$ if he chooses to make a decision $a_j$. If this is linked to my theory, we have to define what is meant by accessible and inaccessible decision variables. Let $\theta$ be accessible if $A$ really is able to perform all the actions $a_1,...,a_n$, and is able to make a decision here. If not, we say that $\theta$ is inaccessible.

To carry out this connection, we have to give meaning to all the Postulates 1 to 4 of Section 3. Postulate 1 gives no problem; all variables connected to $A$ satisfy this postulate. Postulate 2 must be assumed. Then we may consider, corresponding to the concrete decision associated with $\theta$, simpler decisions, with decision variables $\lambda$, such that each $\lambda$ is a function of $\theta$. A way to achieve this is to let these simpler decisions be associated with disjoint subsets of the actions $a_1,...,a_n$. It then seems obvious that the simpler decisions are accessible when the decision connected to $\theta$ is accessible.

Postulate 3 is a challenge here, but it can be satisfied in the following situation: Assume that $A$ has concrete ideals when making his decisions, and he can imagine that one of these ideals has made similar decisions before, but he does not know this so concretely that he can figure out what the ideal person would have done in his concrete case. Let the inaccessible variable $\phi$ correspond to the choices that $A$'s ideal would have done. Alternatively, we can let $\phi$ consist of the whole past history of $A$ together with what may be called the freee will of $A$.

Postulate 4 may be justified by appealing to Zorn's lemma for the partial order defined by taking functions of decision variables. The maximal decisions that can be made by $A$ will have a special place in the proposed quantum decision theory.

If all these assumptions are made, we now have the results of Theorem 6 and Theorem 7, which give a Hilbert space apparatus connected to the situation. We then make the assumption that $A$ really at the same time is confronted with two difficult decisions, each involving decision variables which to him are maximal.

To complete the link to quantum decisions, we must find probabilities connected to the decision variables. For this, one can use the Born formula, which is briefly discussed below; a detailed derivation is given in [1].

All this, and consequences of it, is now discussed in [83].

\subsection{On entanglement and EPR}

Consider two spin 1/2 particles, originally in the state of total spin 0, then separated, one particle sent to Alice and one particle sent to Bob. This can be described by the entangled singlet state
\begin{equation}
|\psi\rangle = \frac{|1+\rangle |2-\rangle - |1-\rangle |2+\rangle}{\sqrt{2}},
\label{spin1}
\end{equation}
where $|1+\rangle$ means that particle 1 has spin component +1 in some fixed direction, and $|1-\rangle$ means that the component is -1; similarly, for $|2+\rangle$ and $|2-\rangle$ with particle 2. 

As in David Bohm's version of the EPR situation, let Alice measure the spin component of her particle in some direction $a$, and let Bob measure the spin component of his particle in the same direction. As has been described in numerous papers, there seemingly is a strange correlation here: The spin components are always opposite.

I want to couple this with the philosophy of Convivial Solipsism [30]: Every description of the world must be relative to some observer. So let us introduce an observer, Charlie, observing the results of both Alice and Bob. Charlie's observations are all connected to the entangled state (\ref{spin1}). 

Let us try to describe all this in terms of accessible and inaccessible variables. The unit spin vectors $\phi_1$ and $\phi_2$ of the two particles are certainly inaccessible to Charlie, but it turns out that their dot product $\eta=\phi_1\cdot \phi_2$ is accessible to him. In fact, Charlie's observations are necessarily related to the state given by $\eta=-3$.

Mathematically this is proved as follows. The eigenvalues of the operator $A^\eta$ corresponding to $\eta$ are 1 and -3. The eigenvector associated with the eigenvalue -3 is just the singlet state $|\psi\rangle$ of (\ref{spin1}), while the eigenspace associated with the eigenvalue 1 is three-dimensional. (See for instance exercise 6.9. page 181 in [48].) So, in the singlet state, we necessarily have $\eta = -3$.

What does it mean that $\eta=\phi_1\cdot \phi_2=-3$? It means that $\phi_{1x}\phi_{2x}+\phi_{1y}\phi_{2y}+\phi_{1z}\phi_{2z}=-3$, and since all components here are either -1 or +1, this is only possible if $\phi_{1x}\phi_{2x}=-1$ etc., which implies $\phi_{1x}=-\phi_{2x}$, $\phi_{1y}=-\phi_{2y}$, and $\phi_{1z}=-\phi_{2z}$. It follows that $\phi_{1a}=-\phi_{2a}$ in every direction $a$. It is assumed here that, even though the derivation is concerned with inaccessible variables, the conclusion, which is an accessible conclusion for some experiment, is nevertheless valid.

Note that Charlie can be any person. So, we conclude: To any observing person, the spin components as measured by Alice and Bob must be opposite. This seems to be a necessary conclusion, implied by the fact that the person, relative to his observations, is related to the state given by (\ref{spin1}).

For the further limitations of the observer Charlie and the corresponding explanation of why the CHSH inequality can be violated in Bell-type experiments, see [34,61,83,90].

\subsection{The Born rule}

Born's formula is the basis for all probability calculations in quantum mechanics, The version given in most textbooks can be formulated as follows: Given some mixed state  $\rho$ the expectation of the theoretical variable $\theta$ is given in terms of the associated operator $A^\theta$ as
\begin{equation}
E(\theta)= \mathrm{trace} (\rho A^\theta).
\label{Born}
\end{equation}

In textbooks, Born's formula is usually stated as a separate axiom, but it has also been argued for by
using various sets of assumptions [54-56]; see also Campanella et al. [50] for some references. In fact, the first argument for the Born formula, assuming that there is an affine mapping from a set of density functions to the corresponding probability functions, is due to von Neumann [51]. Many modern arguments rely on Gleason's theorem, which is not valid in dimension 2, and also requires some assumptions, in particular the assumption of non-contextuality. In [1] a simple version of Born's formula is derived under reasonable assumptions from a Gleason-type theorem due to Busch [52], a theorem which also holds in dimension 2. The assumptions can be formulated in terms of two extra Postulates; see [1,83,88,90].

\subsection{The Hamiltonian and the Schr\"{o}dinger equation}

First: In the approach of this article, I have concentrated on the construction of operators connected to accessible variables. From this, state vectors or ket vectors can in some generality be seen as eigenvectors of some operator; see Section 10 below and also a further discussion in [1].

During a time when no measurement is done on the system, the ket vectors are known in quantum mechanics to develop according to the Schr\"{o}dinger equation:
\begin{equation}
i \frac{h}{2\pi} \frac{d}{dt}|\psi\rangle_t =H |\psi\rangle_t ,
\label{schr}
\end{equation}
where $H$ is a self-adjoint operator, the Hamiltonian. (Referring to the general theory above, this is the operator corresponding to the variable $\theta$= total energy, in a context where the relevant observer or set of communicating observers also think of the complementary variable time).

In [1] I gave two sets of arguments for the Schr\"{o}dinger equation, one rough and general, and then one specific related to position. The last argument also includes
a discussion of the wave function. The general argument for the Schr\"{o}dinger equation will be reproduced here; a wider discussion of time development is given in [83].

\subsection{The general argument for the Schr\"{o}dinger equation; unitary transformations and entanglement}

Assume that the system at time 0 has a state given by the ket $|\psi\rangle_0$ and at time $t$ by the ket $|\psi\rangle_t$. Let's assume that the contexts are given as follows: We can ask an epistemic question about a maximal accessible variable $\theta$, and the ket corresponding to a specific value of this variable is $|\theta\rangle_0$ at time 0
and $|\theta\rangle_t$ at time $t$. We have the choice between making a perfect measurement at time 0 or at time $t$. Since there is no disturbance through measurement
of the system between these two time points, the probability distribution of the answer must be the same whatever choice is made. Hence according to the simple version of Born's formula
\begin{equation}
|_0\langle \theta|\psi\rangle_0 |^2 =|_t\langle \theta|\psi\rangle_t|^2 .
\label{Bequal}
\end{equation}

Now we refer to a general theorem by Wigner [58], proved in detail by Bargmann [59]: If an equation like (\ref{Bequal}) holds, then there must be a unitary or
antiunitary transformation from $|\psi\rangle_0$ to $|\psi\rangle_t$. (Antiunitary $U$ means $U^{-1}=-U^{\dagger}$.) Since by continuity, an antiunitary transformation can
be excluded here, so we have
\[|\psi\rangle_t =U_t|\psi\rangle_0 \]
for some unitary operator $U_t$. Writing $U_t=\mathrm{exp}(\frac{2\pi A_t}{ih})$ for some selfadjoint operator $A_t$, and assuming that $A_t$ is linear in $t$: $A_t=Ht$,
this is equivalent to (\ref{schr}). In fact, assuming that $\{U_t\}$ forms a strongly continuous group of unitary transformations, the form $Ht$ of $A_t$ follows from a theorem by Stone; see Holevo [60].

Unitary transformations of states play an important role in quantum mechanics. Both in the continuous case and in the discrete case such a transformation can be used to illuminate the state
concept as introduced in the present article. More specifically, a unitary change of an operator can quite generally be coupled to a concrete change of the involved theoretical variable; see Theorem 5 and Theorem 7 above. When an operator is changed in this way, its eigenvectors are changed accordingly, hence there is a change of states. Note that, subject to linearity, a unitary operator $U$ always can be written as $U=\mathrm{exp}(\frac{2\pi Ht}{ih})$ for some suitable Hamiltonian $H$, so these transformations
can be seen as closely related to time developments of states.

Consider the discrete case. Let the initial state be $|a;k\rangle \otimes |b;j\rangle$, corresponding to the answers of two focused questions: $\theta^a=u_k^a$ and
$\gamma^b =v_j^b$. Assume that $\theta^a$ and $\gamma^b$ are maximal. By a unitary transformation, essentially by a time development, this initial state is transformed into a state which cannot be written as a product of states
in this way but is a linear combination of such states. This is an entangled state. Thus, in my terminology, entangled states can at least in some cases be given concrete interpretations: Some
fixed time ago they were given as answers to two focused questions. By the inverse unitary transformation, the entangled state may be transformed back to the state
$|a;k\rangle \otimes |b;j\rangle$ again. Thus, we have then a concrete interpretation of the entangled state: Subject to a suitable Hamiltonian, the state can be interpreted as the
answer to two focused questions posed at some past time.

\section{A brief comparison with some other approaches}

As mentioned in the Introduction, there are several rather recent investigations with the purpose of deriving the Hilbert space structure from physical assumptions. Some of the resulting models are called generalized probability models. I will briefly discuss some of these approaches.

The article [6] by Hardy is a pioneering one. This article led to several other investigations, summarized in Hardy [65]. All these investigations start with a set of postulates, stated slightly differently.

Both in [6] and in [65] postulates are stated in terms of two basic numbers $N$ and $K$. Here, $N$ is the maximum number of states for a given system for which there exists some measurement, which can identify which state from a set in a single shot, while $K$ is the number of probabilities that are entries in the state vectors that are constructed. Hardy presents a set of postulates that characterize both classical systems and quantum systems, and one postulate which distinguishes the two. It is proved that $K=N$ for classical systems, and $K=N^2$ for quantum systems. In modern form, Hardy's postulates are stated as P1, P2, P3, P4' and P5 of [65]. These lead to either classical mechanics or quantum mechanics, while a variant, P4, singles out quantum mechanics. In fact, it leads to a very general version of quantum mechanics, which is further discussed in [65].

In contrast, my postulates of Section 3 here can be associated with a version of quantum theory where I limit the pure state concept to ket vectors that are eigenvectors of some physically meaningful operator.

Another difference is that Hardy starts his investigations by introducing pure states and measurements, while I start with the notion of theoretical variables and their associated operators. It is an open problem to find a minimal set of postulates which cover all different approaches towards the formalism. It is crucial that my approach also leads to a view of the interpretation of quantum mechanics, discussed below.

Among the various articles that are related to the ones by Hardy, I can mention Goyal [9], who relies on the framework of information geometry, and Masanes and M\"{u}ller [10], who state 5 axioms based on elementary assumptions regarding preparations, transformations, and measurements,

Rovelli's book [64], which is thought-provoking and very informative, is more concerned with interpretation than with foundation. I find very much in this book that I appreciate, and Rovelli's interpretation of quantum mechanics is close to mine.

\section{Interpretation of quantum mechanics}

Based on the results above, we can now start to discuss the interpretation of quantum theory. The results were based upon accessible theoretical variables $\theta$, which were assumed to be connected to an observer or jointly to a group of communicating observers. From a scientific point of view, these variables are the ones for which questions of the form `What is $\theta$?' or `What will $\theta$ be if we measure it?' can be posed. It is tempting here to cite Rovelli [64]: `I believe that we need to adapt our philosophy to our science,
and not our science to our philosophy.' I fully agree with this. The mathematical discussion above belongs to the domain of science; the philosophical discussion of quantum interpretation should come after this.

My theory seems to be more connected to our knowledge of reality rather than reality itself. I will call this a general epistemic interpretation of quantum mechanics.

There exist several interpretations of quantum mechanics, and the discussions between the supporters of the different interpretations are still going on. In recent years, there have been held a broad range of international conferences on the foundation of quantum mechanics. A great number of interpretations have been proposed; some of them look very peculiar to the layman. The many worlds interpretation assumes that there exist millions or billions of parallel worlds and that a new world appears every time one performs a measurement; there is also a related many mind's interpretation.

On two of these conferences recently there was taken an opinion poll among the participants [14,15]. It turned out to be an astonishing disagreement on many fundamental and simple questions. One of these questions was: Is quantum mechanics a description of the objective world, or is it only a description of how we obtain knowledge about reality? The first of these descriptions is called ontological, and the second is called epistemic. Up to now, most physicists have supported some version of an ontological or realistic interpretation of quantum mechanics, but variants of the epistemic interpretation have received a fresh impetus in recent years.

I look upon my book 'Epistemic Processes' [1] and also this article as a contribution to this debate. An epistemic process can denote any process to achieve knowledge. It can be a statistical investigation or a quantum mechanical measurement, but it can also be a simpler process. The book starts with an informal interpretation of quantum states, which in the traditional theory has a very abstract definition. In my opinion, a quantum state can under wide circumstances be connected to a focused question and a sharp answer to this question, see above.

A related interpretation is QBism, or quantum Bayesianism, see Fuchs [16,17,18] and von Baeyer [19]. (What started as a variant of Bayesianism, has now developed into a somewhat wider interpretation.) The predictions of quantum mechanics involve probabilities, and a QBist interprets these as purely subjective probabilities, attached to a concrete observer.  Many elements in QBism represent something completely new in relation to classical physical theory, in relation to many people's conception of science in general and to earlier interpretations of quantum mechanics. 

QBism has been discussed by several authors. For instance, Herv\'{e} Zwirn's views on QBism, which I largely agree with, are given in [20].

By using group theory, group representation theory, and some simple category theory, I aim to study a general situation involving theoretical variables mathematically, and it seems to appear from this that essential elements of the quantum formulation can be derived under weak conditions. This may be of some scientific relevance. Empirically, it has turned out that the the quantum formalism provides a very extensive description of our world as we know it [21], and in physical situations in microcosmos an all-embracing description.

Focus on the case where the accessible variable $\theta$ takes a discrete set of values. In the case where $\theta$ takes an infinite discrete set of values, we can still prove that Theorem 6 and Theorem 7 hold; the proof goes by taking a limit of cases where $\theta$ takes a finite number of values.

The following simple observation should be noted and is in correspondence with the ordinary textbook interpretation of quantum states: Trivially, every vector $|v\rangle$ is the eigenvector of \emph{some} operators. Assume that there is one such operator $A$ that is physically meaningful, and for which $|v\rangle$ is also a non-degenerate eigenvector, say with a corresponding eigenvalue $u$. Let $\lambda$ be a physical variable associated with $A=A^\lambda$. Then $|v\rangle$ can be interpreted as the question `What is the value of $\lambda$?' along with the definite answer `$\lambda=u$'.

More generally, accepting operators with non-degenerate eigenspaces (corresponding to observables that are accessible, but not maximally accessible), each eigenspace can be interpreted as a question along with an answer to this question.

Binding together these two paragraphs, we can also think of the case where $\lambda$ is a vector, such that each component $\lambda_i$ corresponds to an operator $A_i^{\lambda^i}$, and these operators are mutually commuting. Then $A^\lambda=\bigotimes_i A_i^{\lambda^i}$ has eigenspaces which can be interpreted as a set of questions `What are the values of $\lambda_i\ i=1,2,...$?' together with sharp answers to these questions. In the special case of systems of qubits, H\"{o}hn and Wever [35] have recently proved that there is a one-to-one correspondence between sets of question-and-answer pairs and state vectors.

The following is proved in [1,36] under certain general technical conditions, and also specifically in the case of spin/ angular momentum: Given a vector $|v\rangle$ in a Hilbert space $\mathcal{H}$ and a number $u$, there is at most one pair $(a,j)$ such that $|a;j\rangle=|v\rangle$ modulus a phase factor, and $|a;j\rangle$ is an eigenvector of an operator $A^a$ with eigenvalue $u$. 

The main interpretation in [1] is motivated as follows: Suppose the existence of such a vector $|v\rangle$ with $|v\rangle=|a;j\rangle$ for some $a$ and $j$. Then the fact that the state of the system is $|v\rangle$ means that one has focused on a question (`What is the value of $\lambda^a$?') and obtained the definite answer ($\lambda^a=u$.) The question can be associated with the orthonormal basis  $\{|a;j\rangle ;j=1,2,...,d\}$, equivalently with a resolution of the identity $I=\sum_j |a;j\rangle\langle a;j|$. The general technical result of [1] is also valid in the case where $\lambda^a$ and $u$ are real-valued vectors.

After this, we are left with the problem of determining the exact conditions under which \emph{all} vectors $|v\rangle\in \mathcal{H}$ in the non-degenerate discrete case and all projection operators in the general case can be interpreted as above. This will require a rich index set $\mathcal{A}$ determining the index $a$. This problem will not be considered further here, but this is stated as a general question to the quantum community [36]. But from the evidence above, I will in this paper rely on the assumption that each quantum state/ eigenvector space can be associated in a unique way with a question-and-answer pair. Strictly speaking, this requires a new version of quantum mechanics, where we only permit state vectors that are eigenvectors of some physically meaningful operator.

Superposition of quantum states can be introduced in my setting as follows: Take as a point of departure the states $|a;j\rangle$, each such state interpreted in the way that we know that $\lambda^a=u_j^a$ for a maximally accessible variable $\lambda^a$. Then consider another maximal variable $\lambda^b$ and a hypothetical possible value $u_i^b$ for $\lambda^b$. Since $\sum_j |a;j\rangle\langle a;j|=I$, we have
\begin{equation}
|b;i\rangle=\sum_j |a;j\rangle\langle a;j|b;i\rangle = \sum_j \langle a;j |b;i\rangle |a;j\rangle .
\label{superp}
\end{equation}
Here the corresponding operators $A^a$ and $A^b$ may not commute, and this is a fairly general linear combination of states $|a;j\rangle$. Such linear combinations will then be state vectors. The state $|b;i\rangle$ may be a very hypothetical state, not coupled to the observer's concrete knowledge. Then (\ref{superp}) corresponds to a `do not know' state. 

This discussion of superposition may also be generalized to the double-slit experiment and to more general experiments involving multiple paths; see for instance Rovelli [64] for such experiments. The inference pattern in the double-slit experiment can be explained by a momentum variable in the plane of the slits orthogonal to the two slits, a momentum which by de Broglie's theory is connected to a wave. What is not known, is the position variable in the same direction, in particular, the answer to the question `Which slit?'. The answer to a similar question is also unknown in experiments involving multiple paths.

When $\lambda$ is a continuous scalar or vector variable, we can still interpret the eigenspaces of the operator $A^\lambda$ as questions `What is the value of $\lambda$?' together with answers in terms of intervals or more generally sets for $\lambda$. This is related to the spectral decomposition of $A^\lambda$, which gives the resolution of the identity (compare (\ref{9}) in Appendix 2)
\begin{equation}
I=\int_{\sigma(A^\lambda)} dE(\lambda).
\label{111}
\end{equation}

This resolution of the identity is tightly coupled to the question `What is the value of $\lambda?$', and it implies projections related to indicators of intervals/sets $C$ for $\lambda$, that is, yes-no questions of the type `Does $\lambda$ belong to $C$?', as
\begin{equation}
\Pi (C)=\int_{\sigma(A^\lambda )\cap C} dE(\lambda).
\label{112}
\end{equation}

This is of course just simple quantum logic. But it can be related to an interpretation if we can agree on the basic assumption of Convivial Solipsism: Every description of reality should be relative to an observer or a group of communicating observers. All theoretical variables of this article are assumed to have such a relation. Thus yes-no questions associated with such a variable should be related to an observer or a group of observers.

Then (\ref{112}) can be interpreted as connected to a general epistemic interpretation of quantum states and projection operators. A special case is the QBist interpretation but this interpretation is more general. It can also be seen as a concrete specification of Relational Quantum Mechanics: Accessible variables of a system are seen as relative to other systems, where one of these other systems may be an observer. In the multiple-world interpretation, only variables connected to one world are accessible to a given observer at some fixed time. In the Bohm interpretation, the position of a particle at time $t$ will be accessible, but the path from time $t_1$ to time $t_2$ will be inaccessible.

There is a huge literature on interpretations of quantum theory. Some of the proposed interpretations have relationships to my epistemic interpretation, but I will not go into more details with this discussion here.

In general, $\lambda$ may be seen as a maximal accessible variable associated with the operator $A^\lambda$. If $\theta$ is another maximal accessible variable, it will be associated with another operator $A^\theta$, and $A^\lambda$ and $A^\theta$ will not be commuting. We can then say that $\lambda$ and $\theta$ are complementary variables in the sense of Bohr. More precisely, it is the questions related to these variables that are complementary, each given by an orthonormal full set of ket vectors. Variables/operators corresponding to the same formal question, but having different sharp answers to this question, are equivalent in this respect. They are given by the same orthonormal basis, and the variables are bijective functions of each other.

In a physical context, Niels Bohr's complementarity concept has been thoroughly discussed by Plotnitsky [37]. 

Here is Plotnitsky’s definition of complementarity:

(a)	a mutual exclusivity of certain phenomena, entities, or conceptions; and yet

(b)	 the possibility of applying each one of them separately at any given point; and

(c)	the necessity of using all of them at different moments for a comprehensive account of the totality of phenomena that we consider.

This definition points to the physical situation discussed above and has Niels Bohr’s interpretation of quantum mechanics as a point of departure. In [1] and [83] I have also tried to couple the complementarity concept to macroscopic situations.

Here is one remark concerning QBism, which can be said to represent a special case of my views: Subjective Bayes probabilities have also been in fashion among groups of statisticians. In my opinion, it can be very fruitful to look for analogies between statistical inference theory and quantum mechanics, but then one must look more broadly at statistics and statistical inference theory, not only focusing on subjective Bayesianism. This is only one of several philosophies that can form a basis for statistics as a science. Studying connections between these philosophies is an active research area today. From such discussions, one might infer that another interesting version of Bayesianism is objective Bayesianism, with a prior based on group actions.

Finally, I need to discuss my epistemic interpretation of quantum mechanics in light of the various no-go theorems in the literature.

For the relationship to Bell's theorem, I refer to my recent article [61]. It seems possible to avoid the non-locality assumption if we replace it with an assumption that all observers are limited in some specific sense.

My inaccessible variables are more general than what is usually perceived as hidden variables. Nevertheless, the Kochen-Specker theorem may a priori be of some relevance. However, the Kochen-Specker theorem only excludes noncontextual hidden variable theories. I think of my theory of an observer as connected to a fixed physical context at some fixed time. 

A greater challenge is the recent Pusey-Barrett-Rudolph (PBR) theorem. This theorem seemingly excludes models where a pure quantum state represents only knowledge about an underlying physical state of the relevant system. A crucial assumption, however, is the existence, for every system, of such an underlying `real physical state'. This assumption can be questioned. Also, the arguments against an epistemic interpretation given by Colbeck and Rennes [70] rely on certain assumptions. In particular, it assumes a list of elements of reality $\Lambda$ which satisfies a certain Markov condition. A detailed discussion will not be given here.

It is important, however, that also arguments against a realistic interpretation of quantum states are given in the literature [74,75].

\section{From inaccessible variables to notions}

In the previous Sections, all variables were seen as mathematical variables. In this Section I will limit the variable concept to only physical variables, and these physical variables will be accessible. I will limit the discussion to some concrete physical context, where all physical variables in principle are measurable given some choice by the observer $C$ or by a group of communicating observerss. The space $\Omega_\phi$ will then be replaced by some abstract space $\Omega$, and group theory on $\phi$ will be replaced by category theory with object $\Omega$, called a space of \emph{notions}. It is assumed that these notions exist relative to the observer $C$ or to a group of communicating observers. Functions from $\Omega$ onto $\Omega$ or onto other objects, called morphisms, can be defined. In this setting Postulate 3 can be replaced by:
\bigskip

\textbf{Postulate 3':} \textit{Related to a given physical context there exists a space of notions $\Omega$ such that for each accessible variable $\eta$ there exists a morphism from $\Omega$ onto $\Omega_\eta\subset\Omega$.}
\bigskip

I repeat my definition of a \emph{physical} context: All variables can in principle be measurable by $C$. This can be seen to agree for instance with the context-concept used in the Kochen-Specker theorem.  In [72] contextuality is discussed in more detail using the ordinary formal apparatus of quantum mechanics. This apparatus will not be assumed here. Of some interest for the current approach are various inequalities discussed in [72], in particular Bell inequalities. My explanation of why a particular inequality, the CHSH inequality, has now turned out to be violated in loophole-free experiments, is given in [34,61].

In the previous Sections, I built the arguments on group theory and group representation theory. The basic concepts were theoretical variables and groups acting upon these variables. These variables were assumed to exist relative to an observer or jointly to some communicating observers.

The very concept of a variable assumes something which varies, but in the previous discussion I did not explicitly assume that all variables can take definite values, at least not known to us; see Section 3. To some, this distinction may be problematic. It is not problematic at all for what I have called accessible physical variables. The variable `position' can take values $x_0$, and the variable `momentum' can take values $p_0$. But according to Heisenberg's inequality, we cannot in a physical context assume joint values $(x_0,p_0)$.

In general, we cannot in a physical context assume maximal inaccessible quantities $\phi$ taking physical values. In this Section, I still want to consider an observer or a set of communicating observers, but relative to these, I will in general consider `notions', not only `variables'. A notion is anything that a person can have as a basis for his thinking, or that a communicating set of persons can think jointly of. 

To make this precise, mathematically, we need partly to go from group theory to category theory. Category theory was founded by Mac Lane [62], and has been used by several physicists [63, 65, 69] in the foundation of quantum mechanics. Bob Coecke has in several papers discussed quantum mechanics from the point of view of category theory; see also Hardy  [65].

Group theory is a specialization of the more general category theory, and my task in this Section will be to generalize the discussion of this paper correspondingly. In general terms, a group is a category with one object in which every morphism has a (two-sided) inverse under composition. Just the simplest aspects of category theory will be used in this article.

As said, the space of theoretical variables must be replaced by a space of notions $\Omega$, and transformations in this space must be replaced by labeled arrows, automorphisms, from $\Omega$ onto $\Omega$. The groups $G^a$ associated with accessible variables $\theta^a$, special notions, can be kept as before, but the group $K$ acting on $\phi$ must be replaced by something new. But first, consider the basic postulates in this setting.
\bigskip

\textbf{Postulate 1':} \textit{ It is assumed that if $\theta$ is a notion and there is a morphism defined on $\Omega_\theta\subseteq\Omega$ which maps $\theta$ onto some $\lambda$, then $\lambda$ is a notion.}
\bigskip

Note that everything that is connected to accessible variables goes as before, but the context of the previous Section is now replaced by a physical context, including all variables that are accessible to $C$ before, during and after a measurement. So, in particular, Postulate 2, Definition 1 and Postulate 4 can be taken over from Section 3. Definition 2 must be modified to:
\bigskip

\textbf{Definition 2'} \textit{Let $\theta$ and $\eta$ be two maximal accessible variables in some physical context, and let $v_\theta$ and $v_\eta$ be the morphisms from $\Omega$ onto $\Omega_\theta$ and $\Omega_\eta$, respectively. If there is an automorphism $k$ in $\Omega$ such that $k$ has an inverse and $v_\eta = v_\theta\circ k$, we say that $\theta$ and $\eta$ are related. If no such $k$ can be found, we say that $\theta$ and $\eta$ are non-related.}
\bigskip

This can be seen to be an equivalence relation. Again, it follows from the relationship that $\eta$ is maximal iff $\theta$ is maximal. And if $G$ is a group acting on $\theta$, we can define a group $H$ acting on $\eta$ by $h\circ v_\eta =  (g\circ v_\theta)\circ k$. The mapping from $g$ to $h$ is an isomorphism.

Returning to the two physical examples of Section 3, here is the definition of $k$ in these cases: If $\theta=\theta^a$ and $\eta=\theta^b$ are the spin components in directions $a$ and $b$, we can consider the plane spent by $a$ and $b$, and let $k$ be the $180^\circ$ rotation around the midline between the two vectors. If $\theta$ is the theoretical position and $\eta$ is the theoretical momentum for a particle, we can define $k$ by a Fourier transform.

In Subsection 7.2, it is showed how $k$ can be defined in general in the finite-dimensional case for two variables both  taking $n$ values. In this case, all such pairs of maximal accessible variables are related.

My main basic result is then that Theorem 0 holds also in this setting.
More generally, Theorem 4 and Theorem 6 of Section 7 hold in a physical context with the above four postulates.

Note that the basic requirement is only that we have two different maximal accessible (complementary) variables in the given physical setting. This seems to be a weaker requirement than conditions formulated in other approaches towards the reconstruction of the Hilbert space apparatus. In general, each of the maximal accessible may be a vector, whose components are jointly measurable variables.

Versions of the basic Theorem 5 and Theorem 7 of Section 7 are still valid in this context but will require some more work.
 
Proofs and further discussions are given in Appendix 4.

Again, all the results are valid when the single actor $C$ is replaced by a set of communicating observers, and the notions are shared among these observers.

 There may be an obvious objection against the present approach: Everything is connected to an observer or jointly to a communicating group of observers. How can this be connected to a theory of the real world? My answer to this is that I base everything on the main thesis of Convivial Solipsism: Every description of the world must be from the point of view of an observer or some communicating group of observers. In particular, every accessible variable that $C$ may be thinking of, has a parallel real value variable connected to any measurement of experiment that $C$ may make. Quantum mechanics is a model related to $C$, but in this way, it will also be a model of real experiments and measurements. I say more about the interpretation of quantum mechanics in Section 9.

 \section{Concluding remarks}
 
 The treatment of this paper is not quite complete. Some remaining tasks include:

 - A further development of the case of continuous theoretical variables.
 
 - Giving concrete conditions under which the Born formula is applicable in practice. This is in particular relevant in connection to cognitive modeling.
 
 - Developing an axiomatic basis in the spirit of quantum logic (see for instance [43]). But note the simple postulates of Section 3 above.
 
 - A treatment of open quantum systems.
 
 - A further discussion of the relationship to other approaches and to other interpretations.
 
 - More concrete examples of how this approach can be used to address the conceptual and technical challenges of quantum theory. Some aspects of this are discussed in [47] and in [61].
 
- A discussion of the implications of this approach for the experimental and technological aspects of quantum mechanics.
 \bigskip
 
 Group theory and quantum mechanics are intimately connected, as discussed in detail in several books and papers. In this article, it is shown that the familiar Hilbert space formulation can be derived mathematically from a simple basis of groups acting on theoretical variables. The consequences of this are further discussed in [1,83,90]. The discussion there also seems to provide a link to statistical inference [86,87].
 
 From the viewpoint of purely statistical inference, the accessible variables $\theta$ discussed in this paper are parameters. In many statistical applications it is useful to have a group of actions $G$ defined on the parameter space; see for instance the discussion in [44]. In the present paper, the basic group $G$ is assumed to be transitive, hence, tentatively, if we have a group on some parameter that is not transitive, the quantization of quantum mechanics can be derived from the following principle: all model reductions in some given model should be to an orbit of the group.
 
 It is of some interest that the same criterion can be used to derive the statistical model corresponding to the partial least squares algorithm in chemometrics [45], and this connection also motivates an important case of the recently proposed more general envelope model [46].
 
 The main message of the present paper: the reconstruction of quantum mechanics from simple assumptions, is connected to an observer or jointly to a group of communicating observers, by just taking accessible variables as a primitive notion. But some postulates like Postulate 3 and Postulate 4 seem to be necessary. Then under weak assumptions, the main condition needed seems to be that there exist two different complementary variables, where the word `complementary' is taken to mean different maximal accessible theoretical variables.
 
 In this paper, the first axioms of quantum theory are derived from reasonable assumptions. 
 As briefly stated in [1], one can perhaps expect after this that such a relatively simple theoretical basis for quantum theory may facilitate a further discussion regarding its relationship to relativity theory. One can regard physical variables as theoretical variables, inaccessible inside black holes. These ideas are further developed in [47,84].

 Further aspects of the connection between quantum theory and statistical inference theory, which relies heavily on decisions, decisions that in my view can largely be modeled by using quantum decision theory, are under investigation.
 
 Finally, I want to point out that some of my previous published papers contain certain errors and inaccuracies. In [4] and [5] it is erroneously stated that the basic group representation $U$ should be irreducible. This is now corrected in [4]. Sloppy formulations in [34] are now cleared up in [61]. Most errors are now corrected in the book [66], but the proof there, p. 33, that the variable $\theta$ can be written as a function of the group element $n$, is incorrect. The correct version is in Lemma A2 below. It is a strong hope that all mathematical arguments of the present article are correct.
 
 \section*{Acknowledgments} I am grateful to a reviewer, whose comments led to an improvement of this article. I am also grateful to Richard Gill, who gave me the hint that the maximal inaccessible variable $\phi$ could be avoided by using category theory. Finally, I want to thank Bj\o rn Solheim for some interesting remarks.

\section*{Appendix 1. Proofs of Theorem 4 and Theorem 5}

\subsection*{Basic construction}

In the proof of these theorems in [4], I considered the arbitrary phase of the coherent states involved. Strictly speaking, this is correct but makes the proof more difficult to follow, so I will avoid this subtility here. People who want to be more precise can replace statements of the form $g\in G$ with the corresponding $g\in G/E$, where $E$ is the subgroup generated by the arbitrary phase. See also Perelomov [31], Chapter 2.
\bigskip

So fix a ket vector $|\theta_0\rangle$, and consider the coherent states $U(g)|\theta_0\rangle$, where we may here take as $U(\cdot)$ the left-regular representation $U^L (\cdot)$ defined above Propostion 2 in Section 5. By this Proposition,  these states are in one-to-one correspondence with $\theta$, and we can write $|\theta\rangle = U(g)|\theta_0\rangle$. If the representation $U(\cdot)$ is irreducible, we can refer to the theory of Subsections 6.1 and 6.2. The identity (\ref{6}) holds, the operator $A^\theta$ associated with $\theta$ can be defined by (\ref{7}), and this operator has the properties (i)-(iii) as stated there.

It is crucial for this argument that $U(\cdot)$ is irreducible. It is known that abelian groups only have one-dimensional irreducible representations. So if $G$ is abelian, it is only possible to satisfy (\ref{6}) if $\mathcal{H}$ is one-dimensional, giving a trivial theory.

In the following, I will allow $U(\cdot)$ to be reducible, but maintain the basic construction $|\theta\rangle = U(g)|\theta_0\rangle$.

\subsection*{Two maximal accessible variables}

So, we will stick to the reducible case. For this case, study \emph{two} theoretical variables $\theta$ and $\eta$, whose ranges are in one-to-one correspondence. 

Assume that the variables $\theta$ and $\eta$ are maximal as accessible variables, and that both can be seen as functions of an underlying inaccessible variable $\phi$. Then, perhaps by replacing $\eta$ by a bijective function of the same variable, we may by Proposition 1 of Section 3 assume that there exists a transformation $k$ such that $\eta (\phi)=\theta (k\phi)$. The variables $\theta$ and $\eta$ are related.

Let $g\in G$ be a transitive group action on $\theta$, and let $h\in H$ be the transitive group action on $\eta$  
defined by $h\eta (\phi)=g^1 \theta (k\phi)$ when $\eta(\phi)=\theta(k\phi)$, where $g^1\in G^1$, an independent copy of $G$. This gives a group isomorphism between $G$ and $H$.

Let $n\in N$ be the group actions on $\psi =(\theta ,\eta )$ generated by $G$ and $H$ and a single element $j$ defined by $j\psi =(\eta,\theta)$ and $j\theta =\eta$. For $g\in G$, define $g j\psi (\phi) =(g\theta (k\phi),g\theta(\phi))$ when $\eta =\theta (k\phi)$, and for $h\in H$ define $h j\psi (\phi) =(h\eta (\phi), h \eta (k^{-1}\phi))$ when $\theta(\phi)=\eta (k^{-1}\phi)$. Since $G$ and $H$ are transitive on the components, and since through $j$ one can choose for a group element of $N$ to act first arbitrarily on the first component and then arbitrarily on the second component, $N$ is transitive on $\psi$. Also, $N$ is non-Abelian: $gj\ne jg$.

Consider the representation $U(\cdot)$ of the group corresponding to $G$, acting on some Hilbert space $\mathcal{H}$,   with the property that if we fix some vector $|v_0\rangle\in\mathcal{H}$, then the vectors $U(g)|v_0\rangle$ are in one-to-one correspondence with the group elements $g\in G$ and hence with the values $g\theta_0$ of $\theta$ for some fixed $\theta_0$. I choose to use the notation $|v_0\rangle$ in this proof instead of $|\theta_0\rangle$, since several sets of coherent states will be considered.

For each element $g\in G$ there is an element $h =j g j\in H$ and vice versa. Note that $j\cdot j=e$, the unit element. Let $U(j)=J$ be some unitary operator on $\mathcal{H}$ such that $J\cdot J =I$. Then for the representation $U (\cdot)$ of the group corresponding to $G$, there is a representation $V (\cdot)$ of the group corresponding to $H$ given by $V (j gj)=J U (g)J$. These are acting on the same Hilbert space $\mathcal{H}$ with vectors $|v\rangle$, and they are equivalent in the concrete sense that the groups of operators $\{U(g)\}$ and $\{V(h)\}$ are isomorphic.

Note that $J$ must satisfy $J U(j g j)=U (g)J$.  By Schur's Lemma, this demands $J$ to be an isomorphism or the zero operator if the representation $U(\cdot )$ is irreducible. In the reducible case a non-trivial operator $J$ exists, however:

In such a case there exists at least one proper invariant subrepresentation $U_0$ acting on some vector space $\mathcal{H}_0$, a proper subspace of $\mathcal{H}$, and another proper invariant subrepresentation $U'_0$ acting on an orthogonal vector space $\mathcal{H}'_0$. Fix $|v_0\rangle \in \mathcal{H}_0$ and $|v'_0\rangle \in \mathcal{H}'_0$, and then define $J|v_0\rangle=|v'_0\rangle$, $J|v'_0\rangle=|v_0\rangle$ and if necessary $J|v\rangle=|v\rangle$ for any $|v\rangle\in \mathcal{H}$ which is orthogonal to $|v_0 \rangle$ and $|v'_0 \rangle$.

Now we can define a representation $W(\cdot)$ of the full group $N$ acting on $\psi =(\theta, \eta)$ in the natural way: $W(g)=U(g)$ for $g\in G$, $W(h)=V(h)$ for $h\in H$, $W(j)=J$, and then on products from this.

If $U$ is irreducible, then also $V$ is an irreducible representation of $H$, and we can define operators $A^\theta$ corresponding to $\theta$ and $A^\eta$ corresponding to $\eta$ as in (\ref{7}). If not, we need to show that the representation $W$ of $N$ constructed above is irreducible.
\bigskip

\textbf{Lemma A1} \textit{$W(\cdot)$ as defined above is irreducible.}
\bigskip

\textit{Proof.}
Assume that $W(\cdot)$ is reducible, which implies that both $U(\cdot)$ and $V(\cdot)$ are reducible, i.e., can be defined on a proper sub-space $\mathcal{H}_0\subset\mathcal{H}$, and that $J=W(j)$ also can be defined on this sub-space. Let $R(\cdot)$ be the representation $U(\cdot)$ of $G$ restricted to vectors $|u\rangle$ in $\mathcal{H}$ orthogonal to $\mathcal{H}_0$. Fix some vector $|u_0\rangle$ in this orthogonal space; then consider the coherent vectors in this space given by $R(g)|u_0\rangle$. Note that the vectors orthogonal to $\mathcal{H}_0$ together with the vectors in $\mathcal{H}_0$ span $\mathcal{H}$, and the vectors $U(g)|u_0\rangle$ in $\mathcal{H}$ are in one-to-one correspondence with $\theta$. Then the vectors $R(g)|u_0\rangle$. are in one-to-one correspondence with a subvariable $\theta^1$. And define the representation $S(\cdot)$ of $H$ by $S(jgj) =R(g)$ and vectors $S(h)|v_0\rangle$, where $|v_0\rangle$ is a fixed vector of $\mathcal{H}$, orthogonal to $\mathcal{H}_0$. These are in one-to-one correspondence with a subparameter $\eta^1$ of $\eta$.

Fix $\theta_0 \in \Omega_\theta$. Given a value $\theta$, there is a unique element $g_\theta \in G$ such that $\theta = g_\theta \theta_0$. (It is assumed that the isotropy group of $G$ is trivial.)

From this look at the fixed vector $S(jg_\theta j)|v_0\rangle$. By what has been said above, this corresponds to a unique value $\eta^1$, which is determined by $g_\theta$, and hence by $\theta$. But this means that a specification of $\theta$ determines the vector $(\theta, \eta^1)$, contrary to the assumption that $\theta$ is maximal as an accessible variable.  Thus $W(\cdot)$ cannot be reducible.

$\Box$

Note that it is crucial for this proof that the space $\mathcal{H}$ is multi-dimensional. In particular, the proof does not work for the following case: $\phi = (\theta,\eta)$, the transformation $k$ defining relationship exchanges $\theta$ and $\eta$, and $G$ is just the identity. Then $\mathcal{H}$ would be one-dimensional. If this was allowed in the proof and in the corresponding definition of reducibility, all maximal accessible variables would by definition be related.

\bigskip

This lemma shows that there are group actions $n\in N$ acting on $\psi = (\theta, \eta)$ and an irreducible representation $W(\cdot)$ of $N$ on the Hilbert space $\mathcal{H}$. Hence the identity (\ref{6}) holds if $G$ is replaced by $N$, and the coherent states by $|v_n\rangle = W(n)|v_0\rangle$:

\begin{equation}
\int |v_n\rangle\langle v_n |\nu (dn)=I,
\label{u4}
\end{equation}
where $\nu$ is some left-invariant measure on $N$, and $|v_0\rangle$ is some fixed vector in $\mathcal{H}$.
\bigskip

 \textbf{Lemma A2} \textit{There is a function $f_\theta$ of $n$ such that $\theta=f_\theta (n)$, and a function $f_\eta$ of $n$ such that $\eta=f_\eta (n)$.} 
 \bigskip
 
 \textit{Proof.}
 
 Consider a transformation $n$ transforming $\psi_0 =(\theta_0 ,\eta_0 )$ into $\psi_1 =(\theta_1, \eta_1  )$. There is then a unique $g$ transforming $\theta_0$ into $\theta_1$, and a unique $h$ transforming $\eta_0$ into $\eta_1$. Since the groups $G$ and $H$ are assumed to be transitive and with a trivial isotropy group, the group elements $g$ and $h$ correspond to unique variable elements $\theta$ and $\eta$. These are then determined by $n$.
 
 $\Box$
 
 \bigskip

 We are now ready to define operators corresponding to $\theta$ and $\eta$:

\begin{equation}
A^\theta=\int f_\theta (n)|v_n\rangle\langle v_n |\nu (dn),
\label{u5}
\end{equation}

\begin{equation}
A^\eta=\int f_\eta (n)|v_n\rangle\langle v_n |\nu (dn).
\label{u6}
\end{equation}

These are more precise versions corresponding to equations (17)-(20) in [4].

It is clear that these operators are symmetric when $\theta$ and $\eta$ are real-valued variables. Under a technical assumption, see Proposition A1 below, they will be self-adjoint  Also, if $\theta=1$, then $A^\theta$ is the identity. 

In addition, if $s$ is any transformation in $N$, and $W(\cdot)$ is the representation of $N$ used in the above proof, we have, following the proof of Theorem 3 of Subsection 7.2 and using the left-invariance of $\nu$:

\begin{equation}
W(s^{-1})A^\theta W(s)=\int f_\theta (sn)|v_n\rangle\langle v_n |\nu (dn),
\label{u7}
\end{equation}

Consider an application of this:
\bigskip

Recall that $\theta =\theta(\phi)$, where $\phi$ varies over some space $\Omega_\phi$, and $\phi$ is inaccessible. Let $K$ be some group of transformations of $\Omega_\phi$. Assume that $\theta(\cdot)$ is permissible with respect to $K$, and also that $\eta(\cdot)$ is permissible. Let $T(\cdot)$ be a unitary representation of $K$ such that the coherent states $T(t)|v_0\rangle$ are in one-to-one correspondence with $t$. Then for $t\in K$ the operator $T(t)^\dagger A^\theta T(t)$ is the operator corresponding to $\theta'(\phi)=\theta(t\phi)$.
\bigskip

\textit{Proof}
Since $\theta$ is permissible, $\theta(t\phi)=g(t)\theta(\phi)$ for some transformation $g(t)$ of $\Omega_\theta$. Recall that for $g\in G$, the basic group acting on $\Omega_\theta$, it is assumed that the states $U(g)|v_0\rangle$ are in one-to-one correspondence with $g$. Comparing with the properties of $T(\cdot)$, we must then have $g(t)\in G$, and $T(t)=U(g(t))=W(s(t))$ for an induced transformation $s(t)$ in $N$.

The detailed arguments for this are as follows: $\eta(\cdot)$ is also assumed to be permissible. Hence there is a $h(t)$ such that $\eta(t\phi)=h(t)\eta(\phi)$, so $t$ induces a transformation $s=s(t)$ on $\psi = (\theta,\eta)$. Since $(\theta,\eta)$ is in one-to-one correspondence with $(g,h)$, the transformation $s$ also acts in a unique way on $(g,h)$. By the permissibility of $\theta(\cdot)$,  $g(t)$ must belong to $G$. Since $T(t)$ is in one-to-one correspondence with $t$, and $U(g)$ is in one-to-one correspondence with $g$, we must have $T(t)=U(g(t))$. By the same argument $T(t)=V(h(t))$. Looking at the way $N$ and $W(\cdot)$ are constructed in the above proof, we must have that $s(t)\in N$, and $T(t)=W(s(t))$.

For these transformations and an arbitrary $g\in G$, we can define $\theta'(g)=g(t)\theta(g)$, and get $\theta'(g)=f_\theta (s(t) n_g)$, where $n_g$ is the transformation in $N$ which is induced by $g$ as above. Taking $s=s(t)$ and $W(s)=W(s(t))$ in (\ref{u7}) completes the proof. $\Box$
\bigskip

This also completes the proof of Theorem 5 in Section 7. The first statement there follows from the fact that in this case $j=j(k)$ acts on $\psi=(\theta, \eta)$ and induces a transformation $s(k)$ on the group $N$. Take $s=s(k)$ and $S(k)=W(s(k))$ in (\ref{u7}). Theorem 7 follows as in the main text.

We also have a converse of the first part of Theorem 7. This can be used to make precise the main result of [34,61].
\bigskip

\textbf{Lemma A3} \textit{Consider two maximal accessible theoretical variables $\theta$ and $\eta$. If there is a transformation $s$ of $\Omega_\psi$ such that $A^\eta = W(s)^\dagger A^\theta W(s)$, then $\theta$ and $\eta$ are related.}
\bigskip

\textit{Proof}
By (\ref{u7}) we have
\begin{equation}
A^\eta = \int f_\theta (sn)|v_n\rangle\langle v_n |\mu (dn)
\label{u8}
\end{equation}
for ket vectors $\{|v_n\rangle\}$ for which we also have
\begin{equation}
A^\theta=\int f_\theta (n)|v_n\rangle\langle v_n |\mu (dn).
\label{u9}
\end{equation}
Taking $\theta$ and $\eta$ as the two basic maximal accessible variables in Theorem 4, we also have
\begin{equation}
A^\eta = \int f_\eta (n)|v_n\rangle\langle v_n |\mu (dn)
\label{u10}
\end{equation}
These equations also give a spectral decomposition of $A^\theta$ and $A^\eta$. Since this spectral decomposition is unique, we must have
\begin{equation}
\eta(n) = f_\eta (n)=f_\theta(sn)=\theta(sn),
\label{u11}
\end{equation}
so $\theta$ and $\eta$ are related with respect to to the group $N$, whose elements are in one-to-one correspondence with $\psi=(\theta,\eta )$. Since every such variable by Postulate 3 is a function of $\phi$, they are also related with respect to $\phi$. $\Box$

\section*{Appendix 2. The last part of Theorem 4.}

Recall the following definitions [32] for an operator $A$ with domain $D(A)$ on a Hilbert space $\mathcal{H}$:

- The adjoint $A^\dagger$ of $A$ has domain $D(A^\dagger )$ consisting of every $|u\rangle\in\mathcal{H}$ such that the linear functional $\langle u, A\cdot\rangle$ defined on $D(A)$ is bounded. For $|u\rangle\in D(A^\dagger )$, we define $A^\dagger |u\rangle$ as the unique vector $|v\rangle$ such that $\langle v ,s\rangle = \langle u, A|s\rangle$ for all $|s\rangle \in D(A)$.

- $A$ is self-adjoint if $A^\dagger = A$.

- $A$ is symmetric if $\langle u| Av\rangle = \langle Au|v\rangle$ for all $|u\rangle, |v\rangle\in D(A)$.
\bigskip

\textbf{Lemma A4.}
\textit{The operators $A^\theta$ and $A^\eta$ defined by (\ref{x1}) and (\ref{x2}) in connection to Theorem 4 are symmetric.}
\bigskip

\textit{Proof:}
If $|u\rangle, |v\rangle\in D(A^\theta)$, we have

\begin{equation}
 \langle u|A^\theta v\rangle = \langle A^\theta u|v\rangle = \int f_\theta (n) \langle u|v_n\rangle\langle v_n |v\rangle d\nu (n).
 \label{AA1}
 \end{equation}
 $\Box$
 
 It is easy to see that every self-adjoint operator is symmetric. On the other hand, we have for $A^\theta$ and $A^\eta$:
 \bigskip
 
 \textbf{Proposition A1.}
 \textit{If the integral}
 \begin{equation}
  \int| f_\theta (n)| \langle u|v_n\rangle\langle v_n |u\rangle d\nu (n)
  \label{AA2}
  \end{equation}
  \textit{converges for every $|u\rangle\in D(A^\theta)$, then $A^\theta$ is self-adjoint. A corresponding statement holds for $A^\eta$.}
  \bigskip
  
  \textit{Proof:} See Proposition 3 in [82]. $\Box$
  \bigskip
 
So far, we have found operators for the related maximal variables $\theta$ and $\eta$. Note that these can be any pair of related maximal accessible variables. Now we need to find operators for other accessible variables, maximal or not. I will use Postulate 4: For any accessible variable $\lambda$ there exists a maximal accessible variable $\eta$ such that $\lambda$ is a function of $\eta$. This $\eta$ can now be paired with a $\theta$ which satisfies the assumptions of Theorem 4.

Assume that $\eta$ is real-valued or a real vector, and that we have found a self-adjoint operator $A^\eta$ associated with this $\eta$. Then based on the spectral theorem (e.g., [28]) we have that there exists a projection-valued measure, $E(\eta)$ on $\Omega_\eta$ such that for $|v\rangle\in D(A^\eta)$
\begin{equation}
\langle v|A^\eta |v\rangle =\int_{\sigma(A^\eta)} \eta d\langle v|E(\eta)|v\rangle .
\label{8}
\end{equation}
Here $\sigma(A^\eta)$ is the spectrum of $A^\eta$ as defined in [32].

A more informal way to write (\ref{8}) is

\[A^\eta=\int_{\sigma(A^\eta)}\eta dE(\eta).\]

This defines an orthogonal resolution of the identity
\begin{equation}
\int_{\sigma(A^\eta)}dE(\eta )=I.
\label{9}
\end{equation}

From this, we can define the operator of an arbitrary Borel-measurable function of $\eta$ by
\begin{equation}
A^{f(\eta)}=\int_{\sigma(A^\eta)} f(\eta) dE (\eta).
\label{10}
\end{equation}
In this way, we may by Postulate 4 define operators for all accessible variables, whether they are maximal or not. A special case is when the function $f$ is one-to-one. Then in this way, operators associated with equivalent maximal variables may be defined. This concludes the proof of Theorem 4.

The case with a discrete spectrum is discussed in the main text. In this case, we have
\begin{equation}
A^\eta = \sum_j u_j P_j,
\label{11}
\end{equation}
where $\{u_j\}$ are the eigenvalues and $\{P_j\}$ the projections upon the eigenspaces of $A^\eta$. The equations (\ref{9}) and (\ref{10}) can be written in a similar way.

Important special cases of (\ref{10}) include $f(\eta)=I(\eta\in B)$ for sets $B$, or $f$ can be a linear combination of such indicator variables. As in [1], Subsection 5.3, this can be used to find operators of continuous variables by just using the theory associated with finite-valued variables. This makes the approach towards a theory in Subsection 7.2 especially important.

A further important case is connected to statistical inference theory in the way it is advocated in [1]. Assume that there are data $X$ and a statistical model for these data of the form $P(X\in C|\eta)$ for sets $C$. Then a positive operator-valued measure (POVM) on the data space can be defined by
\begin{equation}
M(C)=\int_{\sigma(A^\eta)} P(X\in C |\eta )dE(\eta).
\label{11}
\end{equation}
The density of $M$ at a point $x$ is called the likelihood effect in [1] and is the basis for the focused likelihood principle formulated there.

Finally, given a probability measure with density $\pi(\eta)$ over the values of $\eta$, one can define a density operator $\sigma$ by
\begin{equation}
\sigma=\int_{\sigma(A^\eta)} \pi(\eta)dE (\eta).
\label{12}
\end{equation}

In [1] the probability measure $\pi$ was assumed to have one out of three possible interpretations: 1) as a Bayesian prior, 2) as a Bayesian posterior or 3) as a frequentist confidence distribution (see [33]).
 
\section*{Appendix 3. Two theorems for the finite-dimensional case}

Consider the case where the maximal accessible variables as in Theorem 6 take a finite number of values. Note that the construction in Subsection 7.1 of an operator corresponding to a variable can be made for any maximal accessible variable $\theta$. If $\theta$ is not maximal, an operator for $\theta$ can be defined by appealing to the spectral theorem. In either case, the operator $A^\theta$ corresponding to $\theta$ has a discrete spectrum. Let the eigenvalues be $\{u_j\}$ and let the corresponding eigenspaces be $\{V_j\}$. The vectors of these eigenspaces are defined as quantum states, and as discussed in the main text, each eigenspace $V_j$ can be associated with a question `What is the value of $\theta$?' together with a definite answer `$\theta =u_j$'. This assumes that the set of values of $\theta$ can be reduced to this set of eigenvalues, which I will justify as follows.
\bigskip

\textbf{Theorem A1}
\textit{Let $\{u_i\}$ be the eigenvalues of the operator $A^\theta$ corresponding to $\theta$. Then it follows that $\Omega_\theta$ is identical to this set of eigenvalues.}
\bigskip

\textit{Proof.}
A proof for the case where $\theta(\cdot)$ is permissible is given in Theorem 4.4, page 94 in [1]. A general proof, using other results in this paper, goes as follows.

Let $\{\theta_i\}$ be the possible values of $\theta$. From (\ref{u5}) we get
\begin{equation}
A^\theta = \sum_i \sum_{j=j(i)}f_{\theta}(n_j ) Q_i = \sum_i \theta_i Q_i ,
\label{13}
\end{equation}
where $\{n_j; j=j(i)\}$ are the elements of the group $N$ such that $\theta_i =f_\theta (n_j)$, and
\begin{equation}
Q_i =r_i \sum_{j=j(i)}|v_{n_j}\rangle\langle v_{n_j}|
\label{14}
\end{equation}
for some constant $r_i$

Consider first the maximal case. Then by Theorem A2 below the eigenvalues of $A^\theta$ are simple, so that we can write
\begin{equation}
A^\theta = \sum u_i |u_i\rangle\langle u_i |,
\label{15}
\end{equation}
where $u_i$ and $|u_i\rangle$ are the different eigenvalues and orthogonal eigenvectors of $A^\theta$. We have to prove that there is some connection between (\ref{13}) and (\ref{15}) in this case.

Assume that one value of $\theta$, say $\theta_1$, is an eigenvalue of $A^\theta$. The other values of $\theta$ are then given by $\theta_i = g_i \theta_1$, where $g_i$ is any member of the group $G$, which as in Subsection 7.2 can be taken to be the cyclic group.

In (\ref{14}) we have $|v_{n_j} \rangle = W(n_j)|v_0\rangle=U(g_i)|v_0 \rangle$, which implies that $U(g_{i'})Q_i U(g_{i'})^\dagger$ for $i'\ne i$ is equal to some other $Q_{i''}$. It follows from $A^\theta =\sum_i \theta_i Q_i$ that 1) $U(g_{i'} )A^\theta U(g_{i'})^\dagger = A^\theta$, 2) If $\theta_1 =u_1$ is an eigenvalue, then we must have that $\theta_i =g_i u_1$ is an eigenvalue for all $i$, since a cyclic permutation of $\{u_i\}$ leaves (\ref{15}) invariant, and a cyclic permutation of $\{\theta_i\}$ leaves (\ref{13}) invariant.

Let $I_0 =\{u_j : u_j =g\theta_1\ \mathrm{for\ some}\ g\in G\}$. Since $G$ is transitive on $\Omega_\theta$, it follows that $I_0 =\Omega_\theta$.

Above, I have assumed that one value of $\theta$, $\theta=\theta_0$ was an eigenvalue of $A^\theta$. So, the conclusion so far is that if one value is an eigenvalue, then all values in $\Omega_\theta$ are eigenvalues. Now the same arguments could have been used with respect to the operator $B=\gamma A^\theta$ for some fixed constant $\gamma\ne 0$. For each $\gamma$ the conclusion is: Either (i) all values in $\Omega_\theta$ are eigenvalues of $B$, or (ii) no values in
$\Omega_\theta$ are eigenvalues of $B$.

Now go back to the general definition (\ref{x1}) of $A^\theta$. Changing from $A^\theta$ to $B$ here, amounts to changing $\theta$ to $\theta'=\gamma\theta$. It is clear that we always can choose $\gamma$ in such a way that there is one value in $\Omega_{\theta'}$ which equals the first eigenvalue of $B$. Thus, the conclusion (i) holds for one choice of $\gamma$. Now the change from $\theta$ to $\theta'$ also changes the measure $\mu$ which is involved in the definition of the operator and also in a corresponding resolution (\ref{x0}) of the identity. It is only one choice of $\gamma$, namely $\gamma=1$ which makes the resolution of the identity (\ref{x0}) valid, which is crucial for the theory. Thus, one is forced to conclude that $\gamma=1$, and that the conclusion (i) holds for this choice.

Hence $\Omega_\theta$ is contained in the set of eigenvalues of $A^\theta$. If there were one eigenvalue that is not contained in  $\Omega_\theta$, one can use this eigenvalue as a basis for choosing $\gamma$ in the argument above, hence getting a contradiction. Thus, the two sets are identical.

Having proved this for a maximal accessible $\theta$, it is clear that it also follows for a more general accessible $\lambda =f(\theta)$, since the spectrum then is changed as in (\ref{10}).

$\Box$.
\bigskip

We also have the following:
\bigskip

\textbf{Theorem A2} 
\textit{The accessible variable $\theta$ is maximal if and only if each eigenspace $V_j$ of the operator $A^\theta$ is one-dimensional.}
\bigskip

\textit{Proof.}
The assertion that there exists an eigenspace that is not one-dimensional, is equivalent with the following: Some eigenvalue $u_j$ correspond to at least two orthogonal eigenvectors $|j\rangle$ and $|i\rangle$. Based on the spectral theorem, the operator $A^\theta$ corresponding to $\theta$ can be written as $\sum_r u_r P_r$, where $P_r$ is the projection upon the eigenspace $V_r$. Now define a new e-variable $\psi$ whose operator $B$ has the following properties: If $r\ne j$, the eigenvalues and eigenspaces of $B$ are equal to those of $A^\theta$. If $r=j$, $B$ has two different eigenvalues on the two one-dimensional spaces spanned by $|j\rangle$ and $|i\rangle$, respectively, otherwise its eventual eigenvalues are equal to $u_j$ in the space $V_j$. Then $\theta=\theta(\psi)$, and $\psi\ne\theta$ is inaccessible if and only if $\theta$ is maximal accessible. This construction is impossible if and only if all eigenspaces are one-dimensional.
$\Box$

\section*{Appendix 4. Proofs related to the category theory approach}

In these proofs, I will use the following assumption: For any accessible variable $\theta$, if $v_\theta$ is a morphism from $\Omega$ onto $\Omega_\theta$, then there exists at least one set of elements $u\in\Omega$ such that $\theta=v_\theta (u)$. It is an open question whether the proofs can be generalized to hold without this assumption.
\bigskip

\underline{Proof of Theorem 4 and Theorem 6.} Define first the group $H$ acting on $\eta$ by $h\circ\nu_\eta = (g\circ\nu_\theta)\circ k$. Since this implies  $g\circ\nu_\theta = (h\circ\nu_\eta)\circ k^{-1}$, the groups $G$ and $H$ are isomorphic.

The next step is to define a new group $N$ acting on $\psi = (\theta,\eta)$. To this end, we first define $g\psi =(g\theta,\eta)$ and $h\psi =(\theta, h\eta)$ and then a mapping $j$ from $\Omega_\psi$ to $\Omega_\psi$ by $j\psi=(\eta, \theta )$. 

For $g\in G$, define $(g j)\psi (u) =(((g\circ \nu_\theta)\circ k)(u),(g\circ \nu_\theta)(u))$ when $\nu_\eta(u) =(\nu_\theta\circ k)(u)$, and for $h\in H$ define $(h j)\psi (u) =(((h\circ \nu_\eta)\circ k^{-1})(u),(h\circ \nu_\eta)(u))$ when $\nu_\theta(u) =(\nu_\eta\circ k^{-1})(u)$. It follows that $h=jgj$.

The representation $W(\cdot)$ of $N$ is defined as before, and the proof of irreducibility of $W$ goes as before. Also, the proof of Lemma A2 can be carried out as before. From this, operators $A^\theta$ and $A^\eta$ are again defined by (\ref{u5}) and (\ref{u6}), and operators for other accessible variables are found by using the spectral theorem.

The proof of Theorem 6 goes as in Subsection 7.2.
$\Box$
\bigskip

The proof of Theorem 5/ Theorem 7 will require some more work. One seemingly needs to replace the representation theory for groups with a more general theory. This can be done by considering the multiplicative group $M$ of square matrices over complex numbers, such that a space $K$ of automorphisms $\Omega\rightarrow \Omega$ are mapped into $M$, and these mappings are homomorphisms. Such a set of homomorphism will be called a representation $T$ of $K$. If the matrices in the domain of $T$ are unitary, we say that the representation is unitary. 

However, such a general theory will not be assumed below. I will limit the discussion to the case where $K$ is a group acting on the elements $u\in\Omega$. Recall first Definition 2'  from Section 10.

What we do need here, is a generalization of the concept of permissibility.
\bigskip

\textbf{Definition 4} \textit{Consider a space $K$ of automorphisms $\Omega\rightarrow\Omega$. An accessible variable $\theta$ is said to be permissible with respect to $K$ if there exists a morphism $v$ from $\Omega$ onto $\Omega_\theta$ with the following properties: 1) For any automorphism $t$ of $\Omega$ and any $k\in K$, the identity $v\circ t =v$ implies $v\circ(k\circ t)=v\circ k$. 2) If $k \in K$ is fixed and $(v\circ k)\circ (v\circ k')=(v\circ k)\circ (v\circ k'')$ or $(v\circ k')\circ (v\circ k)=(v\circ k'')\circ (v\circ k)$ are identities in $\Omega_\theta$, then $v\circ k' =v\circ k''$.}
\bigskip

\textbf{Lemma 3} \textit{ If $K$ is a group and $\theta$ is permissible with respect to $K$ under the morphism $v$, then $G=v\circ K$ is a group.}
\bigskip

\underline{Proof.} Check that under permissibility every $g\in G$ is an automorphism on $\Omega_\theta$ with a two-sided inverse. $\Box$
\bigskip

\textbf{Theorem 5'} \textit{Assume that the related accessible variables $\theta$ and $\eta$ are permissible with respect to a group $K$ of automorphisms in $\Omega$, under the morphisms $v_\theta$ and $v_\eta$. Assume that $K$ is transitive and has a trivial isotropy group. Let $T$ be a unitary representation of $K$ such that the coherent states $T(t)|\psi_0\rangle$ are in one-to-one correspondence with $t\in K$. For any transformation $t\in K$ and any such representation $T$ the operator $T(t)^\dagger A^\theta T(t)$ is the operator of $\theta'$, the image from $\Omega$ under the morphism $v_\theta\circ t$.}

\textit{In addition, since $\theta$ and $\eta$ are related, there is a unitary operator $W$ such that $A^\eta =W^\dagger A^\theta W$.}
\bigskip

\underline{Proof.} For the first part use a modification of the proof of Theorem 5 in Appendix 1. First, by permissibility and Lemma 3, we have that for given $t\in K$ there exist group elements $g(t)=v_\theta \circ t$ and $h(t)=v_\eta\circ t$ such that $\theta (t\circ u)=g(t)\theta(u)$ and $\eta (t\circ u )=h(t)\eta (u)$ for any $u\in \Omega$. This induces a transformation $s=s(t)$ on $\psi=(\theta,\eta)$. The rest of the proof goes as before.

For the second part, use (\ref{u7}) and the fact that the righthand side of (\ref{u7}) is equal to (\ref{u6}). (see the proof of Lemma A3.)  $\Box$
\bigskip

For the finite-dimensional case, we can define $k$ and $K$ acting on $\psi=(\theta,\eta)$ as in Subsection 7.2. Theorem 7 seems then to be true also in this setting as it is formulated in that Subsection with the modifications that Postulate 1' to Postulate 4 of Section 4 are assumed, and $\theta'(t)$ is defined as $v_\theta(t\circ u)$ for $u\in\Omega$ such that $\theta=v_\theta (u)$.

\end{document}